\documentclass[a4paper,11pt]{article}
\pdfoutput=1 

\usepackage{jheppub} 

\usepackage[T1]{fontenc} 
\usepackage{blindtext}
\usepackage{epstopdf}
\usepackage{graphicx}
\usepackage{epsfig}
\usepackage{dcolumn}  
\usepackage{bm}    
\usepackage{caption}
\usepackage{subcaption}
\usepackage{amssymb} 
\usepackage{tikz}
\usetikzlibrary{arrows}
\usepackage{nccmath}
\usepackage{xcolor}
\usepackage{epstopdf}
\usepackage{graphicx,wrapfig,lipsum}
\usepackage{epsfig}
\usepackage{amsmath,bm}
\usepackage{amsfonts}  
\usepackage{amsmath}  
\usepackage{slashed}  
\usepackage{enumitem}
\usepackage[mathscr]{euscript}
\usepackage{tabu}
\usepackage{epsfig}
\DeclareMathOperator{\arccosh}{arcCosh}
\DeclareMathOperator{\arcsech}{arcsech}

\makeatletter
\g@addto@macro\bfseries{\boldmath}
\makeatother

\def\l1{{{1-loop}}}

\def\n1{\Bigg|_{n=1}}

\def\n{{(n)}}

\usepackage[T1]{fontenc} 
\usepackage{tikz}
\usepackage{amsmath,amssymb}
\usepackage{relsize}
\usepackage{latexsym}
\usepackage{leftidx}
\usepackage{xcolor}
\usepackage{diagbox}
\usepackage[T1]{fontenc}
\usepackage{array}
\usepackage{makecell}
\usepackage{csquotes}
\usepackage{tikz}
\usepackage{enumitem}
\usepackage{setspace}
\usepackage{multirow}
\usepackage{amsmath,amssymb}
\usepackage{relsize}
\usepackage{latexsym}
\usepackage{leftidx}
\usepackage{xcolor}
\usepackage{csquotes}
\usepackage{tikz}
\usetikzlibrary{decorations.pathmorphing}
\usetikzlibrary{arrows.meta}
\usepackage{enumitem}
\usetikzlibrary{decorations.markings}
\usetikzlibrary{decorations.pathmorphing}
\usetikzlibrary{decorations.markings}
\usetikzlibrary{decorations.pathmorphing}
\usepackage{pifont}
\usepackage{hyperref}
\usepackage{url}

\usepackage{bookmark}

 \title{\textbf{\textsf{Thermal one point functions, large $d$   and interior geometry of black holes
 }}}
  \author{Justin R. David, Srijan Kumar}
\affiliation{\vspace{.1cm} Centre for High Energy Physics, \\ Indian Institute of Science,\\
C. V. Raman Avenue, Bangalore 560012, India.}
\emailAdd{justin@iisc.ac.in, srijankumar@iisc.ac.in}

\abstract{We study thermal one point functions of massive  scalars 
 in $AdS_{d+1}$ black holes. These are induced by coupling  the scalar to 
  either  the Weyl tensor squared or the Gauss-Bonnet term.
Grinberg and Maldacena argued that the one point functions  sourced by the 
Weyl tensor exponentiate in the limit of large  scalar masses  and 
 they contain information  of the black hole geometry behind the horizon. 
 We observe that the one point functions behave identically in this limit for either of the couplings mentioned earlier. 
 We show that in an appropriate large $d$  limit,  the one point function for the charged black hole 
 in $AdS_{d+1}$ can be obtained  exactly. These black holes in general contain  an inner horizon. 
We  show that the one point 
 function exponentiates and  contains the information of both the proper time between the outer horizon
 to the inner horizon as well as the proper length from the inner horizon to the singularity. 
 We also show that  Gauss-Bonnet coupling induced one point functions  in $AdS_{d+1}$ black holes with hyperbolic 
 horizons behave as anticipated by Grinberg-Maldacena. 
 Finally, we study the one point functions in the background of rotating BTZ black holes induced by the cubic coupling 
 of scalars. 
}

\begin{document} 
\maketitle
\flushbottom
\section{Introduction}

The physics of the horizon of black holes and the geometry behind the horizon have continued to 
be problems of interest in quantum gravity. 
The AdS/CFT correspondence  \cite{Maldacena:1997re}
in principle gives us a handle to study the geometry behind the horizon 
using   boundary correlators of the CFT at finite temperature. 
Studies in this direction were initiated some time ago  by approximating 2-point functions of 
massive scalars  in terms of 
space like geodesics in the large mass limit \cite{Fidkowski:2003nf}. 
Recently Grinberg and Maldacena \cite{Grinberg:2020fdj} have argued in general that
even one point functions of massive scalars 
in $AdS_{d+1}$ black hole geometries  contain some information 
of the geometry behind the horizon.  
 This has been investigated further in 
\cite{Rodriguez-Gomez:2021pfh,McInnes:2022pig,Georgiou:2022ekc,Berenstein:2022nlj}. 

One point functions in holographic backgrounds have been  studied earlier to establish sum-rules and high 
frequency behaviour of transport coefficients \cite{David:2011hy,David:2012cd,Myers:2016wsu}.
Using an example studied by  \cite{Myers:2016wsu},  Grinberg and Maldacena observe the thermal one point function 
in the planar $AdS_{d+1}$ Schwarzschild black hole evaluated for scalars dual 
to operators of low dimensions can be analytically continued to large operator dimensions. 
When this is done, the thermal one point function exponentiates and one can read out 
the proper time to singularity from the event horizon. 
The one point function is sourced by the coupling of the scalar to the Weyl tensor squared term.
Using the intuition from this  exactly solvable example, \cite{Grinberg:2020fdj} developed a 
saddle point approximation for large masses. The saddle involved geodesics at complex 
radial positions and arguments justifying the location of the saddle and the contour involved in 
the WKB approximation. 
One of the interesting claims of their saddle point analysis is that for the $AdS_5$ charged black hole 
which has an inner horizon, the one point function exponentiates to the form
\begin{equation}\label{chrgexp}
	\langle {\cal O} \rangle  \sim \exp ( - m ( \ell_{\rm hor} + \ell_{\rm sing } - i \tau_{\rm in }) ) \ .
\end{equation}
Here $\tau_{\rm in}$ is the proper time for a particle to reach the inner horizon from the outer horizon, 
$\ell_{\rm hor}$ is the regularized  proper length from the boundary to the outer horizon and 
$\ell_{\rm sign}$ is the proper length from the inner horizon to the singularity. 
The lengths are shown on the Penrose diagram of the charged black hole in figure \ref{3lenghts}. 
In this analysis, the massive scalar field is taken to be a probe and its  the back reaction  is neglected. 
One of the aims of this paper is to provide a solvable  example similar to the   $AdS_{d+1}$
Schwarzschild black hole to show indeed that the expectation value of the thermal one point functions
in the charged black hole indeed exponentiates as in (\ref{chrgexp}).

In this paper we study one point functions of massive scalars sourced by the Gauss-Bonnet curvature 
as well as the Weyl tensor squared. 
For the  solvable case of planar Schwarzschild  black hole we see that the result for the thermal one point function 
sourced by the Gauss-Bonnet curvature is identical to that of the Weyl tensor squared term. 
We then present
examples in which the one point function of massive scalars can be exactly obtained
for small conformal dimensions just as the example seen first in \cite{Myers:2016wsu}
for the $AdS_{d+1}$ Schwarzschild black hole. 
The first example is the planar charged black hole in $AdS_{d+1}$ in a suitable large 
$d$ limit.  Large $d$ limit has been studied earlier by several groups, see  \cite{Emparan:2020inr} for 
a recent review.  The large $d$ limit used in this paper also involves a simultaneous limit on the charge 
of the black hole, which is distinct from that discussed in \cite{Emparan:2020inr} \footnote{See the recent work 
of \cite{Giataganas:2021jbj} for another instance in which holographic observables are exactly solvable in a
large $d$ limit.}.
We see that the one point function can be obtained exactly and it exponentiates for large scalar masses just as expected by 
arguments in \cite{Grinberg:2020fdj}. 
We show that the one can read out all the $3$ lengths in (\ref{chrgexp}) from the one point function. 
We also observe that this result is independent of whether the one point function is sourced by 
Gauss-Bonnet curvature or the Weyl tensor squared. 

The second example we study are the black holes with hyperbolic horizons in $AdS_{d+1}$. 
Here we show that the Gauss-Bonnet curvature sources one point function which again behave as anticipated
by \cite{Grinberg:2020fdj}. 
We can read out the $\tau_{s}$, the proper time from the horizon to the singularity from the one point function.
The dependence of $\tau_{s}$ on the $AdS$ radius is different from Schwarzschild black holes with 
planar horizons. 
Therefore this example serves another check of the general arguments of \cite{Grinberg:2020fdj}.

Finally we study the one point functions in the background of the rotating BTZ black hole. 
By conformal invariance, thermal one point functions of primaries vanish in $2d$ CFT's on a line.
However if the spatial directions are periodic, they acquire non-trivial expectation values. 
Holographically these one point functions are sourced due to cubic couplings
of the massive scalar with other scalars in the theory \cite{Kraus:2016nwo}. 
The auxiliary scalar sources the scalar of interest due to the non-trivial windings around the circular horizon 
of the BTZ. 
These one point functions were explored for the non-rotating BTZ in  \cite{Grinberg:2020fdj}. As expected, they decay exponentially 
as $\exp\big( - \frac{2\pi L}{\beta}\big)$, were $L$ is 
 size of the spatial identifications.  However, from the coefficient of this leading term, we can extract the  proper
 time to singularity. 
 We generalize this computation to the rotating BTZ and obtain the exact one point function sourced due to the cubic coupling of 
 scalars. 
 We then take a limit which retains the information of the inner horizon,  it is possible to interpret the coefficient 
 of  $\exp\big( - \frac{2\pi L}{\beta}\big)$ geometrically. 
 From this coefficient, we see that we can read out both $\tau_s$ and $\ell_{\rm hor}$, but there is no 
 dependence on $\ell_{\rm sing}$. We verify that this behaviour is also exhibited by the expectation value of 
 composite operators $\langle {\cal O}^2 \rangle$ of large integer scaling dimensions in finite temperature  2d CFT. 
 The WKB type arguments used in \cite{Grinberg:2020fdj} assumed radial symmetry. 
 It will be interesting to generalize those arguments to the case of rotating black holes in higher dimensions and as well try to obtain the
 thermal one 
 point functions exactly to see if they behave as in (\ref{chrgexp}).

 The organization of the paper is as follows. 
 In  section \ref{section2},  we evaluate the one point function  of massive scalars  in the planar  Schwarzschild black hole 
 in $AdS_{d+1}$ 
 sourced by 
  the Gauss-Bonnet coupling, we observe that  the result for the one point function 
 is identical to that of the Weyl-tensor squared coupling. 
 In section \ref{section3}, we  repeat the analysis  for charged 
 planar black holes in $AdS_{d+1}$ in a suitable large $d$ limit. 
 Section \ref{hyperbhsec} contains the evaluation of the one point function in hyperbolic black holes. 
 In section \ref{section4} we study the thermal  one point function sourced by cubic coupling of scalars 
 in the rotating BTZ black hole. 
 Section \ref{section5}  has our conclusions. 
 Appendix \ref{appendixa}  has details regarding the evaluation of the Greens functions needed to evaluate the 
 thermal one point function. 
 Appendix \ref{appendixb} discusses the details of the planar charged  black hole  in $AdS_{d+1}$ in the large
 $d$ limit developed in the paper. 
 Appendix  \ref{appendixc} shows that the result for the one point function sourced  by the Weyl tensor squared 
 coupling for the charged planar black hole 
  in the large mass limit is the same as that obtained due to the Gauss-Bonnet coupling.

\section{Gauss-Bonnet induced thermal one point function}
\label{section2}

In this section we will re-visit the set-up  of \cite{Grinberg:2020fdj} 
which results in   non-trivial thermal  one point functions  due to the presence of a  higher derivative coupling in the 
low energy effective action.  
Consider the minimally coupled scalar $\varphi$ of mass $m$ in an  asymptotically 
$AdS_{d+1}$ background. 
The scalar is dual to an operator ${\cal O}$ of dimensions  \cite{Witten:1998qj}
\begin{equation}
\Delta = \frac{d}{2} + \sqrt{ \frac{d^2}{4} + m^2R^2_{AdS} } \,\, ,
\end{equation}
where $R_{AdS}$ is the radius of $AdS$. 
If one considers the conventional  quadratic  action of $\varphi$,  the symmetry $\varphi\rightarrow - \varphi$
prevents the dual operator ${\cal O}$ developing an expectation value. 
In \cite{Grinberg:2020fdj}, the following 
 the low energy effective action of the scalar was considered
\begin{equation}
S = \frac{1}{16\pi G_N} \int  \sqrt{g} d^{d+1} x 
\left[ \frac{1}{2} \left( \nabla^\mu  \varphi \nabla_\mu  \varphi  + m^2 \varphi^2 \right) 
+ \alpha \varphi W_{\mu\nu\rho\sigma} W^{\mu\nu\rho\sigma}  \right]  ,
\end{equation}
where the scalar $\varphi$ couples to  the Weyl tensor squared. 
Such couplings are known to be present in the low energy effective action 
 and are dual to the 3-point function involving $2$ stress tensors and 
the operator ${\cal O}$.   Since  it is a  higher derivative coupling,  it is suppressed in the strong t'Hooft coupling limit.
 From dimensional analysis 
it is clear that 
\begin{equation}
\alpha \sim l_s^2 \sim \frac{R_{AdS} ^2 }{\sqrt \lambda } \, ,
\end{equation}
where $l_s$  is the string length and   $\lambda$ is the  t' Hooft coupling and the 
gauge theory. 
One of the motivations provided in \cite{Grinberg:2020fdj}  for this choice of  the coupling is that the Weyl tensor vanishes 
for the pure $AdS$, which is consistent with the one point function of the operator ${\cal O}$ vanishing at 
zero temperature. 

It has been observed that in the study of black hole entropy the Gauss-Bonnet coupling 
results in the same corrections to black hole entropy as the Weyl tensor squared \cite{Sen:2005iz}. 
Indeed low energy effective actions in string theory contain a coupling of the Gauss-Bonnet term with 
the dilaton.  Motivated by these observations, we will evaluate the expectation value of the one 
point function by considering the action
\begin{equation} \label{gbaction}
S = \frac{1}{16\pi G_N} \int  \sqrt{g} d^{d+1} x 
\left[ \frac{1}{2} \left( \nabla^\mu  \varphi \nabla_\mu  \varphi  + m^2 \varphi^2 \right) 
+ \alpha \varphi {\cal L}_{\rm GB}  \right] \, ,
\end{equation} 
where the Gauss-Bonnet term is given by 
\begin{equation}
{\cal L}_{\rm GB}  = R_{\mu\nu\rho\sigma} R^{\mu\nu\rho\sigma} - 4 R_{\mu\nu} R^{\mu\nu} + R^2  .
\end{equation}
We will show that the result for the one point function  is identical to that obtained by \cite{Grinberg:2020fdj}. 
Though the Gauss-Bonnet  curvature for pure AdS is non-vanishing, the one point function  evaluated 
using the coupling  in (\ref{gbaction}) 
does vanish for the pure $AdS$. 
Therefore the result   is consistent with the vanishing of thermal expectation values at zero temperature. 

Using the rules of $AdS/CFT$, the one-point function is given by 
\begin{equation} \label{1ptgen}
\langle O(t, \vec x ) \rangle  = \alpha \int dz' dt' d\vec x'
 \sqrt{g} \tilde K( t, \vec x; \,  z', t', \vec x') {\cal L}_{\rm GB} (z', t', \vec x' ) \, ,
\end{equation}
\footnote{$d\vec x'= (dx)^{d-1}$} where   $ \tilde K(t, \vec x;\,  z', t', \vec x' ) $ is the bulk-boundary propagator.
The  integral over $ t', \vec x' $
is carried over  
 the $AdS_{d+1}$ Schwarzschild black hole. 
 More specifically, the range of integration for the $t', x'$ coordinates runs from 
 $-\infty$ to  $\infty$ and the integral over $z'$ runs from the horizon to infinity. 
  The metric of the  planar
 Schwarzschild black hole is  given by 
	\begin{align} \label{adsbh}
		ds^2=\frac{R^2}{z^2}\bigg(-f(z)dt^2+\frac{dz^2}{f(z)}+d\vec{x}^2\bigg),\ \ \ \ \ f(z)=1-\frac{z^d}{z_0^d}\ ,\ \ \ \ z_0=\frac{\beta d}{4\pi} \, .
	\end{align}
Here $R$ is the radius of $AdS_{d+1}$\footnote{From this point onwards $R$ will refer to the radius of $AdS$}
 and $\beta$ is the inverse temperature. 
The  Gauss-Bonnet curvature of the geometry in (\ref{adsbh}) is given by 
\begin{equation}\label{gbadsbh}
{\cal L}_{\rm GB} (z)  =d (d-2) ( d-1)^2  \frac{ z^{2d}}{ z_0^{2d} R^4} + 
\frac{( d-2)( d-1)d (d+1) }{R^4} \, .
\end{equation}
It is also useful to evaluate the Weyl tensor squared
\begin{equation}
W_{\mu\nu\rho\sigma}W^{\mu\nu\rho\sigma}  (z)  =d (d-2) ( d-1)^2  \frac{ z^{2d}}{ z_0^{2d} R^4} \, .
\end{equation}
Thus  the Gauss-Bonnet term and the Weyl tensor squared term differs by a constant. 
We will see that this additional constant in (\ref{gbadsbh}) does not contribute to the one point function. 

We follow the strategy of \cite{Grinberg:2020fdj} 
to obtain  the bulk-boundary propagator. 
We first solve for the bulk-bulk Greens function  in the geometry (\ref{adsbh}) by solving the equation
\begin{equation}
\frac{1}{\sqrt{-g} } \partial_\mu( \sqrt{-g} g^{\mu\nu}  \partial_\nu G( z, z', t,t', x, x')  - m^2  G( z, z', t,t', x,x') 
= \frac{\delta (z-z') \delta (t-t') \delta^{d-1} ( \vec x-\vec x') }{ \sqrt{-g}} \, .
\end{equation}
Then the bulk-boundary propagator can be obtained by  \cite{Banks:1998dd,Erbin}
\begin{equation} \label{bbp}
K (z', t,t', x,x') = \lim_{z\rightarrow 0 } \frac{2\nu}{ z^\Delta} G( z, z', t, t', x, x')  \, .
\end{equation}
Since the  geometry  in (\ref{adsbh}) has translation invariance in $t, \vec x$ directions, we can expand the Greens function 
in terms of the Fourier modes in these directions \footnote{For the Euclidean black hole $t$ is periodic, therefore the 
Fourier modes in this direction are discrete, this is understood in the paper. For convenience we will denote the sum over 
these modes by the integral over $\omega$. }
\begin{equation}
G(z, z', t, t', x, x') = \int \frac{d\omega d\vec k }{ ( 2\pi )^d} e^{i \omega (t-t') - \vec k \cdot ( \vec x- \vec x') }  
\hat G( z, z', \omega, \vec k ) \, .
\end{equation}
The differential equation obeyed by the Fourier coefficient is given by 
\begin{eqnarray}
&& \partial_z\left( \frac{R^{d-1}}{z^{d-1} } f(z)  \partial_z \hat G(z, z', \omega, \vec k ) \right)  - \frac{R^{d+1}}{z^{d+1}} m^2 \hat G(z, z', \omega, \vec k )  \\ \nonumber
&& \qquad\qquad\qquad\qquad\qquad + \frac{R^{d-1}}{z^{d-1}} 
\left( \frac{\omega^2}{f(z) } -  \vec k^2 \right) \hat G(z, z' \omega,\vec k )  = 
\delta( z-z') \, .
\end{eqnarray}
Therefore the bulk-boundary propagator also admits the expansion 
\begin{equation}
K(z', t,t', x,x') =  \int \frac{d\omega d \vec k }{ ( 2\pi )^d} e^{i \omega (t-t') - \vec k ( \vec x- \vec x') }  
\hat K( z, z', \omega, \vec k ) \, .
\end{equation}
This expansion can then be used in (\ref{1ptgen}) to obtain  Fourier components of the expectation value,
which can be written as 
\begin{equation}
\langle{\cal  O} \rangle_{\omega, \vec k } = \int \sqrt{-g} dz' dt' dx' \hat K ( z', \omega, \vec k ) e^{i (\omega t' - \vec k \cdot \vec x' )} 
{\cal L}_{GB} ( z', t', \vec x') \,.
\end{equation}
From (\ref{gbadsbh}) we see that the Gauss-Bonnet curvature only depends on the $z$ coordinate, 
This  allows us to perform the $t', x'$ integral to obtain 
\begin{equation}
\langle {\cal O} \rangle_{\omega, \vec k }  = ( 2\pi)^d \delta (\omega) \delta ( \vec k ) 
\int dz' K( z', 0, 0) {\cal L}_{GB} ( z') \, .
\end{equation}
Thus the only non-zero Fourier coefficient of the expectation value is its zero mode. 
This implies that the thermal one point functions $\langle {\cal O} ( t , \vec x ) \rangle$ is uniform, independent of time and position.
It is convenient to define the zero mode of the bulk-boundary Green's function by 
\begin{equation}
\hat K( z') = \hat K(z', 0, 0) \ .
\end{equation}
From (\ref{bbp})  we see that 
\begin{equation} \label{bulklim}
\hat K ( z') = \lim_{z\rightarrow 0} \frac{2\nu}{ z^{ \Delta} }  \hat G(z, z') \ ,  
\qquad {\rm where, } \qquad \hat G(z, z')  = \hat G(z, z', \omega , \vec k )|_{\omega = \vec k =0} \ \,,
\end{equation}
and $\hat G(z, z')$ satisfies the differential equation
\begin{equation}
\partial_z\left( \frac{R^{d-1}}{z^{d-1} } f(z)  \partial_z G(z, z') \right)  - \frac{R^{d+1}}{z^{d+1}} m^2 G(z, z')  = 
\delta( z-z') \,.
\end{equation}
Finally the uniform value of the  thermal expectation value is given by 
\begin{equation}\label{1ptfnf}
\langle  {\cal O}  \rangle =   \alpha R^{d+1}\int_{0}^{z_0} \frac{ dz}{  z^{( d+1)} }   K(z)  {\cal L}_{\rm GB} (z) \,.
\end{equation}
Though this result is obvious because of the translational invariance of the geometry (\ref{adsbh}), we have gone through the steps in detail so that we can generalise this discussion when we consider hyperbolic black holes in 
section \ref{hyperbhsec}.

The Green's function $G(z, z')$  has to be regular both at $z=0$ and at $z=1$. 
The details of evaluation of this Greens function is given in the appendix \ref{appendixa}.
Let
\begin{equation}
w = \left( \frac{z}{z_0}\right)^d,  \qquad\qquad\qquad h = \frac{\Delta}{d} \,, 
\end{equation}
then 
\begin{eqnarray}
	G(w,w')&=&-\frac{\Gamma(h)^2}{\Gamma(2h)}\ \frac{z_0^d}{R^{d-1}d}\bigg(\varphi_{{\rm inf} }(w)\varphi_{{\rm hor} }(w')\theta(w'-w)+\varphi_{{\rm hor} }(w)\varphi_{{\rm inf} }(w')\theta(w-w')\bigg) , \nonumber \\
	\varphi_{{\rm inf} }(w)&=&w^h\ {}_2F_1(h,h,2h,w) \,, \qquad  \varphi_{{\rm hor} }(w)= w^h {}_2F_1(h,h,1,1-w)\,.
	\end{eqnarray}
Since the hypergeometric function admits a taylor series expansion around the origin, we see that 
$g_{{\rm inf} }(w)$ is well behaved at  infinity $w=0$, while $g_{{\rm hor}}$ is well behaved at the horizon $w =1$. 
We can now use (\ref{bulklim}) to obtain the bulk to boundary Greens function
	\begin{align}
	K(w)
	=-\frac{\Gamma(h)^2}{\Gamma(2h)}\ \frac{z_0^d}{R^{d-1}d}\ \bigg(\frac{2\nu }{z_0^{\Delta_+}}\bigg) w^h\ {}_2F_1(h,h,1,1-w) \,.
	\end{align}
	Substituting this in the expression for the one point function in (\ref{1ptfnf}) we obtain 
	\begin{eqnarray} \label{1ptfnr}
	\langle O\rangle&=& -\alpha \frac{2\nu}{ d z_0^\Delta R^2}  \frac{\Gamma(h)^2}{\Gamma(2h)} 
	\int_0^1 dw \left[  (d-2)( d-1)^2  w^h \ {}_2F_1(h,h,1,1-w)  \right.  \nonumber \\ 
	& & \qquad\qquad\qquad\qquad \left. +  (d-2) ( d-1) ( d+1)  w^{h-2}   \ {}_2F_1(h,h,1,1-w) \right]  ,  \nonumber\\
	&=& \alpha\frac{(d-2)(d-1)^2}{ R^2 d} \frac{\Gamma(h)^2}{\Gamma(2h)}\bigg(\frac{2\nu }{z_0^{\Delta_+}}\bigg)\ h(h-1)\pi\csc(h\pi) \,.
	\end{eqnarray}
	Here the second term in the integrand arising from the constant term in the Gauss-Bonnet curvature given in 
	(\ref{gbadsbh}). 
	To obtain the last line we have used the result 
	\begin{eqnarray}\label{intident}
		\int_0^1 \, _2F_1(h,h;1;w) (1-w)^{h+n} \, dw&=&\frac{\Gamma (-h+n+2) \Gamma (h+n+1)}{\Gamma (n+2)^2}
		\\ \nonumber
		&& \text{ if } {\rm Re} (h-n)<2\ \text{and}\ {\rm Re} (h+n)>-1 \,.
		\end{eqnarray}
		Note that the integral vanishes for $n=-2$, therefore the constant contribution of the Gauss-Bonnet curvature
		vanishes.  This ensures the result for the one point function from the Gauss-Bonnet coupling is identical to that of the Weyl tensor squared coupling. 
		
Having obtained the one point in (\ref{1ptfnr}), to extract the time to singularity, we take the large $h$ limit with $h\rightarrow \infty - i \epsilon$, this results in 
\begin{eqnarray}  \label{schwexpo}
\lim_{h \rightarrow \infty - i\epsilon } \langle {\cal O} \rangle  &\sim& e^{ - i \pi h }  2^{ - 2h} , \\\nonumber
&=& e^{ - i \pi \frac{m R}{d} }  4^{ - \frac{ m R}{d} } .
\end{eqnarray}
Here we have ignored all the polynomial terms in $h$ and retained only terms which are exponential in $h$. 
We use the fact that ${\rm Im }\;  h <0$ to pick up the negative phase, but we can very well pick up the positive phase
when ${\rm Im } \; h > 0$.

\subsection*{Comparison with geometric lengths}

\begin{figure}[h]
		\centering
		\includegraphics[scale=0.6]{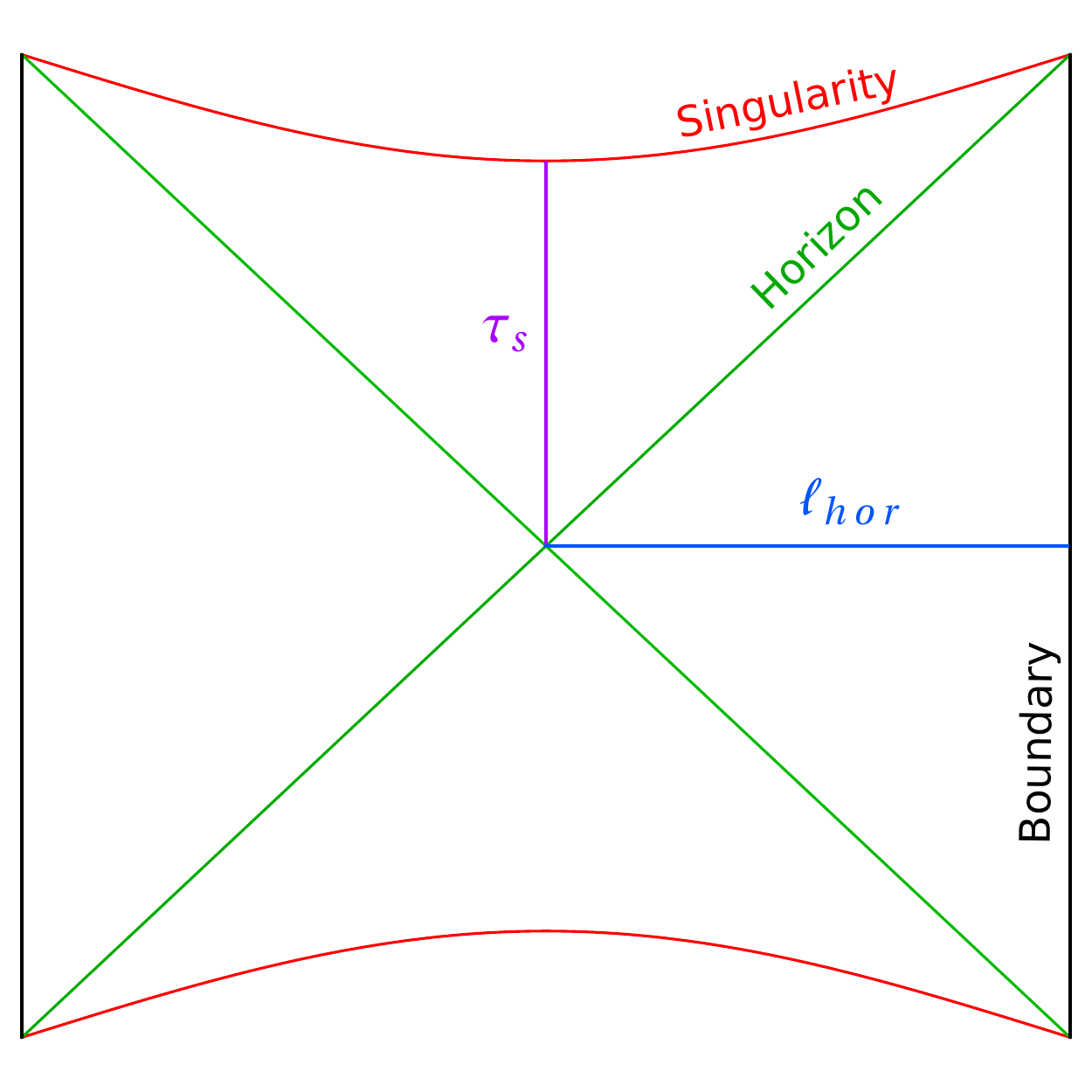}
		\caption{Penrose diagram for the AdS-Schwarzschild metric: $ \tau_s $ is the time-like distance between the horizon and the singularity, and $ \ell_{\rm hor} $ refers to the regularised space-like distance between the boundary and the horizon.} 
		\label{pensch}
	\end{figure}

Maldacena and Grinberg  \cite{Grinberg:2020fdj}, identified the exponential terms occurring in the expectation value of the one point function 
with the following geometric lengths associated with the black hole. 
Consider a radially infalling geodesic with zero energy released at $z=z_0$, the horizon. 
The proper time, the particle takes 
to reach the singularity $z= \infty$ is given by the integral 
\begin{eqnarray} \label{lsing}
t_{s} &=& \int_{z_0}^\infty \frac{R dz}{ z \sqrt{ \big( \frac{z}{z_0}  \big)^d - 1} } , \\ \nonumber 
&=& \frac{\pi R}{d} .
\end{eqnarray}
Similarly, the regularized space like length from infinity to the horizon is given by 
\begin{eqnarray} \label{cutoffdef}
\hat \ell_{{\rm hor}} &=&  \lim_{\epsilon\rightarrow 0}  \int_{\epsilon}^{z_0}  \frac{R dz}{ z \sqrt{1-  \big( \frac{ z}{z_0} \big)^d } },
\\ \nonumber
&=&\frac{R}{d} \log 4 - R \log \big( \frac{\epsilon}{z_0} \big)  + O(\epsilon) .
\end{eqnarray}
The regularized length is obtained by ignoring the divergence as normalised in \cite{Grinberg:2020fdj},  which results in 
\begin{equation} \label{lhor}
\ell_{\rm hor} =  \frac{R}{d}  \log 4 \,.
\end{equation}
One point worth mentioning at this stage is that both  lengths are independent of the mass of the   black hole. 
The mass unlike these lengths can be  obtained  easily by studying the behaviour of particles in the black hole geometry. 
Comparing (\ref{lsing}) and  (\ref{lhor}), we see that in the $h\rightarrow\infty$ limit we can re-write the thermal 
expectation value in the following geometric form
\begin{equation}
\lim_{m \rightarrow \infty - i\epsilon } \langle {\cal O} \rangle \sim 
\exp ( - i m  \tau_{s} - m \ell_{\rm hor} ) \,.
\end{equation}
These lengths are shown on the Penrose diagram of the $AdS$ Schwarzschild black hole in figure \ref{pensch}.

%

\section{The charged planar black hole at large $d$}
\label{section3}

\begin{figure}\centering
		\begin{tikzpicture}[scale=1.5]
			\draw (0,0)--(0,3);
			\draw[dashed] (0,0)--(3,3);
			\draw (3,0)--(3,3);
			\draw[dashed] (3,0)--(0,3);
			\draw[dashed] (3,3)--(0,6);	
			\draw[dashed] (0,3)--(3,6);
			\draw[decorate,decoration={snake},color=red]	(0,3)--(0,6);
			\draw[decorate,decoration={snake},color=red]	(3,6)--(3,3);
			\draw[color=blue] (1.5,4.5)--(3,4.5) node[above,xshift=-.75cm] {$ \ell_{sing} $};
			\draw[color=blue] (1.5,1.5)--(1.5,4.5) node[left,yshift=-1.94cm] {$ \tau_{\rm in} $};
			\draw[color=blue] (1.5,1.5)--(3,1.5) node[below,xshift=-.75cm] {$ \ell_{\rm hor} $};
			\draw[Stealth-] (2.05,2)--(3.5,2) node[right] {Outer horizon};
			\draw[Stealth-] (2.05,4)--(3.5,4) node[right] {Inner horizon};
			\draw[Stealth-] (3,.9)--(3.5,.9) node[right] {Boundary};
			\draw[Stealth-,color=red] (3.05,5)--(3.5,5) node[right] {Singularity};
		\end{tikzpicture}
		\caption{The three lengths $ \tau_{\rm in},\ell_{\rm hor} ,  \ell_{\rm sing} $ shown on the Penrose diagram of the 
		charged black hole in $AdS$. }
	\label{3lenghts}
	\end{figure}
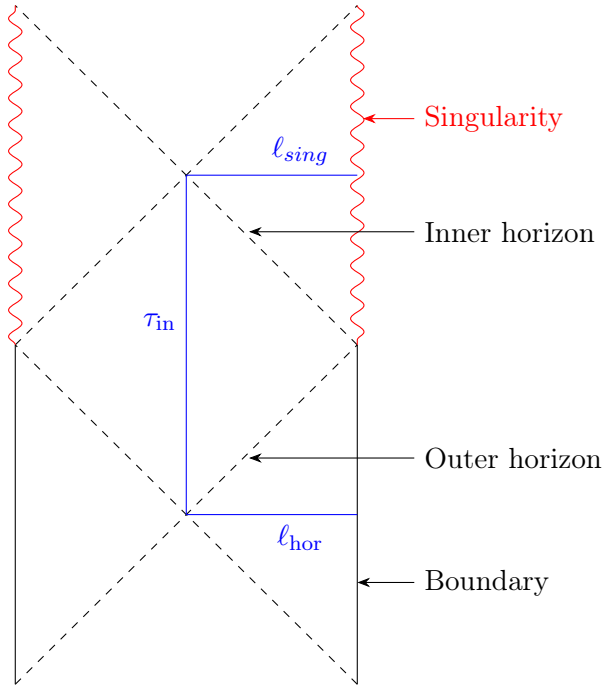

Let us consider the planar Reissner-N\"{o}rdstrom black hole in $AdS_{d+1}$ whose metric is given by 
\cite{Chamblin:1999tk}. 
\begin{align} \label{chargmet}
	ds^2=\frac{R^2}{z^2}\bigg(-f(z)dt^2+\frac{dz^2}{f(z)}+d\vec{x}^2\bigg),\ \ \ \ \ f(z)=1-\frac{z^d}{z_0^d}+q^2z^{2d-2} .
	\end{align}
	In this section, we wish to evaluate the one point function of the scalar in this background  with the action 
	(\ref{gbaction}). 
	Maldacena and Grinberg \cite{Grinberg:2020fdj} deal with the case of charged black hole in $AdS_{5}$. 
	Using a saddle point analysis in terms of  complex geodesics,  they argue that the one point function of the 
	operator dual to a minimally coupled scalar is of the form
	\begin{equation}
	\langle {\cal O} \rangle  \sim \exp ( - m ( \ell_{\rm hor} + \ell_{\rm sing } - i \tau_{\rm in }) )\, ,
	\end{equation}
	where $\ell_{\rm hor}$ is the regularized  distance from the boundary to the outer horizon, 
	$\ell_{\rm sing}$ is the distance from the inner horizon to the singularity and finally 
	$\tau_{\rm in }$ is the  time from the inner horizon to the outer horizon. 
	These lengths are shown in the figure \ref{3lenghts} . 
	Note that here compared to the case without the inner horizon there is an  additional dependence of 
	$\ell_{\rm sing}$. 
	The argument in \cite{Grinberg:2020fdj} involved 
	 a saddle with a contour that goes 
	 over to   complex radial positions.  It would be more satisfying if  the direct analysis similar to that performed in 
	for the Schwarzschild black hole in section \ref{section2} can be done for a black hole with inner horizon. 
	In this section we show that indeed it is possible to arrive at an exact expression for the thermal one point function 
	in a suitable large $d$ limit. 
	Just as in the Schwarzschild case in (\ref{1ptfnr}), the integral which results in the one point function involves 
	an integral from the outer horizon to infinity. 
	
	To begin, let us examine the equation of the minimally coupled scalar  (\ref{chargmet}) 
	in the background  for the zero mode
	in the directions along the boundary. 
	\begin{eqnarray} \label{mcsc}
	&& (1-w)w^2\varphi''(w)-w^2\phi'(w)-h(h-1)\varphi(w) \\ \nonumber
	&& \qquad\qquad\qquad\qquad\qquad
	+\frac{q^2w^{3-\frac{2}{d}}z_0^{2d-2}}{d}\bigg(2(d-1)\varphi'(w)+wd\varphi''(w)\bigg)=0 \, .
	\end{eqnarray}
	Here we have used the co-ordinate
	\begin{equation} \label{defw}
	w =  \left( \frac{z}{z_0} \right)^d.
	\end{equation}
	We will see that $w$ is always finite in the domain we wish to evaluate the integral involving the 
	one point function. 
	To obtain a more tractable equation, we take the following large $d$ limit 
	\begin{equation} \label{dinfty}
	d\rightarrow \infty, \quad {\rm with} \qquad   q^2z_0^{2d-2} = Q, \qquad {\rm held\; fixed}.
	\end{equation}
	In this limit the equation (\ref{mcsc}) \footnote{In appendix \ref{appendixb}, we discuss how 
	 taking the large $d$ limit as in (\ref{dinfty}), 
	 the  solution in (\ref{chargmet})  satisfies the leading order Einstein's equation.} 
	reduces to 
	\begin{align} \label{hypereq}
	(1-w)w^2\varphi''(w)-w^2\varphi'(w)-h(h-1)\varphi(w)+Qw^{3}[2\varphi'(w)+w\varphi''(w)]=0 \, .
	\end{align}
	Solutions of this differential equation can be written in terms of hypergeometric functions. 
	Before we go ahead,  let us obtain the locations of the horizons in the limit (\ref{dinfty}). 
	The function $f(z)$ in  (\ref{chargmet}) becomes 
	\begin{eqnarray}
	f(w) = 1 - w + Q w^2 w^{-\frac{2}{d}} , \\  \nonumber
	\hat f(w) = \lim_{d\rightarrow\infty }  f(w)  = 1 - w + Q w^2 \,.
	\end{eqnarray}
	Therefore the inner and the outer horizons are  at 
	\begin{equation}
	w_{\rm in}  =  \frac{1 + \sqrt{ 1- 4Q}}{2Q} , \qquad w_{\rm out} = \frac{ 1- \sqrt{ 1- 4Q} }{2Q} \,.
	\end{equation}
	
	To construct the Green's function we need the well behaved solutions of (\ref{hypereq})  
	at infinity and at the horizon. 
	These are given by 
	\begin{eqnarray}\label{infhor}
	\varphi_{\rm inf} = w^h (w-2-\sqrt{1-4Q}\ w)^{-h}\ {}_2F_1(h,h,2h,\frac{2\sqrt{1-4Q}\  w}{2+(\sqrt{1-4Q}-1)w}), 
	\\ \nonumber
	\varphi_{\rm hor} = w^h (w-2-\sqrt{1-4Q}\ w)^{-h}\ {}_2F_1(h,h,1,1-\frac{2\sqrt{1-4Q}\  w}{2+(\sqrt{1-4Q}-1)w})\, .
	\end{eqnarray}
	It is easy to observe that $\varphi_{\rm inf}$  is well behaved at the boundary $w\rightarrow 0$, while 
	$\varphi_{\rm hor}$ is well behaved at the outer horizon $w_{\rm out}$. 
	The Greens function satisfies the equation
	\begin{equation}
	\partial_w ( \hat f (w)  \partial_w G(z, z') ) - \frac{h (h-1) }{w^{2} } = \frac{z_0^d}{ d R^{d-1}} \delta ( w-w')  \,.
	\end{equation}
	Here note that the function $f (w) $ has been replaced by its limit $\hat f(w)$. 
The Green's function is obtained by using the two solutions in (\ref{infhor}) and is given by 
\begin{eqnarray}
	G(w,w') &=& -2^{2h}(1-4Q)^{h-\frac{1}{2}}\frac{(\Gamma(h))^2}{\Gamma(2h)}\frac{z_0^d}{R^{d-1} d} \\ \nonumber
	&& \qquad \times \bigg(\varphi_{\rm inf} (w)\varphi_{\rm hor} (w')\theta(w'-w)+\varphi_{\rm hor} (w)\varphi_{\rm inf} (w')\theta(w-w')\bigg) \,.
	\end{eqnarray}
	The bulk boundary propagator using (\ref{bulklim}) is given by 
	\begin{eqnarray} \label{bbpropc}
	K(w)  &=& -2^{2h}(1-4Q)^{h-\frac{1}{2}}\frac{(\Gamma(h))^2}{\Gamma(2h)}\frac{z_0^d}{R^{d-1} d} (-2)^h 
	\left( \frac{2\nu}{z_0^\Delta} \right)  \\
	\nonumber 
	&& \quad\times w^h (w-2-\sqrt{1-4Q}\ w)^{-h}\ {}_2F_1(h,h,1,1-\frac{2\sqrt{1-4Q}\  w}{2+(\sqrt{1-4Q}-1)w}) \,.
	\end{eqnarray}
Finally, we need the limiting value of the Gauss-Bonnet curvature in the large $d$ limit. 
Evaluating the Gauss-Bonnet curvature for the geometry in (\ref{chargmet}), we obtain 
\begin{eqnarray}
	&&\mathcal{L}_{GB}=\frac{(d-2) (d-1) }{R^4}\times\\ \nonumber
	& &\Big((3 d-5) (3 d-4) Q^2 w^{4-\frac{4}{d}} 
	-2 Q \big[d \{d (4 w-1)-10 w+5\}+6 (w-1)\big] w^{ 2- \frac{2}{d}} \\ \nonumber
	 & &\qquad\qquad\qquad+d \left[(d-1) w^2+d+1\right] \Big) \,.
	\end{eqnarray}
	Taking the limit $ d\rightarrow\infty $,  the leading contribution to the Gauss Bonnet term 
	is given by 
	\begin{align}\label{lgb}
	\lim_{d\rightarrow \infty}
	\mathcal{L}_{GB}= \frac{d^4}{R^4} \left( 
	9  Q^2 w^4- 8 w^3  Q+ (2 Q+1) w^2 + 1\right) \,.
	\end{align}
	
	We now have all the necessary  ingredients to evaluate the one point function in the large $d$ limit. 
	Substituting the leading contribution of the Gauss-Bonnet curvature (\ref{lgb}) and  the bulk boundary propagator  (\ref{bbpropc}) 
	in (\ref{1ptfnf}), we obtain  for the thermal one point function
		\begin{align} \label{rati1pt}
		\begin{split}
	\langle {\cal O} \rangle=-& \alpha 2^{2h}\frac{d^2(-2)^{-h}(1-4Q)^{h-\frac{1}{2}}}{R^2}\frac{\Gamma(h)^2}{\Gamma(2h)}\bigg(\frac{2\nu }{z_0^{\Delta_+}}\bigg)\int_{0}^{w_{\rm out} }dw \left[{9  Q^2 w^2}-{8 w  Q}+{ (2 Q+1) }+\frac{1}{w^2}\right]\\
	&\times w^h (w-2-\sqrt{1-4Q}\ w)^{-h}\ {}_2F_1(h,h,1,1-\frac{2\sqrt{1-4Q}\  w}{2+(\sqrt{1-4Q}-1)w}) \,.
	\end{split}\end{align}
	One simple check of this result is to note that when $Q\rightarrow 0$, this expression reduces to the 
	integral obtained in the uncharged  planar black hole in (\ref{1ptfnr}) with $d\rightarrow \infty$. 
	Examining (\ref{lgb}) and (\ref{rati1pt}), we see that we need to keep the ratio $\frac{d}{R} $ to be finite in the large $d$ limit
	to ensure these quantities are finite. 
	Note that the integral is from the outer horizon to infinity, therefore in the domain of integration in 
	 $w$ is always finite.  We can bring the range of integration from $0$ to $1$ by making the substitution
	 \begin{equation}\label{changeofvari}
	 y=1-\frac{2\sqrt{1-4Q}\  w}{2+(\sqrt{1-4Q}-1)w} \,,
	 \end{equation}
	 The one point function then is given by 
		\begin{align} \label{1ptcharge}
		\begin{split}
	\langle{\cal  O} \rangle=&\frac{4 d^2(1-4Q)^{h/2}}{R^2}\frac{(\Gamma(h))^2}{\Gamma(2h)}\bigg(\frac{2\nu }{z_0^{\Delta}}\bigg) \int_{0}^{1}dy\ \, _2F_1(h,h;1;y)\times
	\\&\left[-\frac{9 (1-y)^{h+2}}{4 Q^2 w_{\rm in} ^4 (1-\chi  y)^4}+\frac{2 (1-y)^{h+1}}{Q^2 w_{\rm in}^3 (1-\chi  y)^3}-\frac{(2 Q+1) (1-y)^h}{4 Q^2 w_{\rm in}^2 (1-\chi  y)^2}-\frac{(1-y)^{h-2}}{4}\right] ,
	\end{split}\end{align}
	where
	\begin{equation}
	\chi = \frac{ 1- \sqrt{ 1- 4Q}}{ 1+ \sqrt{1- 4Q} } \,.
	\end{equation}
	The integral of the first three terms in the square bracket in (\ref{1ptcharge})  can be done using 
	formula (7.512.9) in \cite{gradshteyn2007} \footnote{\begin{align*}\begin{split}
		\int_0^1dx\ x^{\gamma-1}(1-x)^{\rho-1}(1-\chi x)^{-\sigma}\ _2F_1(\alpha,\beta,\gamma,x)=&\frac{(1-\chi )^{-\sigma } (\Gamma (\gamma ) \Gamma (\rho ) \Gamma (-\alpha -\beta +\gamma +\rho ))}{\Gamma (-\alpha +\gamma +\rho ) \Gamma (-\beta +\gamma +\rho )}\\
		&\times \, _3F_2\left(\rho ,\sigma ,-\alpha -\beta +\gamma +\rho ;-\alpha +\gamma +\rho ,-\beta +\gamma +\rho ;\frac{\chi }{\chi -1}\right).
		\end{split}\end{align*}}. 
		From (\ref{intident}), we can see that integral of the last term in the square bracket
		vanishes. 
		Using the result for the integrals we obtain
			\begin{align} \label{finachargr}
			\begin{split}
	&\langle{\cal  O} \rangle=-\frac{ d^2(1-4Q)^{\frac{h}{2}-1}}{2R^2 (1-\sqrt{1-4 Q})}\frac{(\Gamma(h))^2}{\Gamma(2h)}\bigg(\frac{2\nu }{z_0^{\Delta}}\bigg) {\pi  (h-1) h \csc (\pi  h)}\times\\
	&\bigg\{\left[\left(-3 h^2-h+10\right) Q+\sqrt{1-4 Q} \left((3 h^2-3 h-2) Q+2\right)-2\right]\times  \\
	& \qquad\qquad\qquad{} _2F_1\left(2-h,h+1;2;\frac{1}{2}-\frac{1}{2 \sqrt{1-4 Q}}\right)\\
	&\hspace{4cm}+4Q (h-2)  \, _2F_1\left(3-h,h+1;2;\frac{1}{2}-\frac{1}{2 \sqrt{1-4 Q}}\right)\bigg\} \,.
	\end{split}\end{align}
	Here the ${}_3F_2$ hypergeometric function simplifies to ${}_2F_1$ for the values of the 
	parameters resulting from the integral. 

Having obtained the leading contribution to the one point function in the large $d$ limit 
we can analytically continue $h$ and take the large $h$ limit. 
Note that 
\begin{equation}
h = \frac{\Delta}{d} \,.
\end{equation}
As we are taking the  large $d$ limit, we would need to ensure that $\Delta$ grows faster than linear in $d$ so that 
$h$ remains large.  $h$ occurs as a parameter of the hypergeometric functions in (\ref{finachargr}), therefore we need the 
asymptotic expansion for large orders. 
To proceed, we first write down the asymptotic expansion obtained by Watson \cite{Watson},  for hypergeometric
functions at large orders
\begin{align}\begin{split}
	{}_2F_1(\alpha+\lambda,\beta-\lambda,\gamma;\frac{1}{2}-\frac{z}{2})\sim\frac{\Gamma(1-\beta+\lambda)\Gamma(\gamma)}{\pi\Gamma(\gamma-\beta+\lambda)}2^{\alpha+\beta-1}(1-e^{-\zeta})^{\frac{1}{2}-\gamma}(1+e^{-\zeta})^{\gamma-\alpha-\beta-\frac{1}{2}}\\
	\times\sum_{s=0}^{\infty}\bigg[c_se^{(\lambda-\beta)\zeta}\Gamma(s+\frac{1}{2})\lambda^{-s-\frac{1}{2}}+c_s'e^{\mp\pi i(\frac{1}{2}-\gamma)}e^{-(\lambda+\alpha)\zeta}\Gamma(s+\frac{1}{2})\lambda^{-s-\frac{1}{2}}\bigg].
	\end{split}\end{align}
	This expansion is valid for large $|\lambda|$ with $\alpha, \beta, z$ fixed, the constants $c_s$ can be found in 
	\cite{Watson}.  Here
	\begin{equation}
	\zeta  = \arccosh z
	\end{equation}
	For our purpose it is sufficient to obtain, the leading term in this expansion. So we can retain the first 
	term with $s=0$. 
	All other terms are sub-leading, they are either exponentially or polynomially suppressed 
	in $\lambda$.  Therefore we have
	\begin{align}\begin{split}
	& {}_2F_1(\alpha+\lambda,\beta-\lambda,\gamma;\frac{1}{2}-\frac{z}{2}) \\
	& \qquad \sim\frac{\Gamma(1-\beta+\lambda)\Gamma(\gamma)}{\pi\Gamma(\gamma-\beta+\lambda)}2^{\alpha+\beta-1}(1-e^{-\zeta})^{\frac{1}{2}-\gamma}(1+e^{-\zeta})^{\gamma-\alpha-\beta-\frac{1}{2}}
	e^{(\lambda-\beta)\zeta}\Gamma(\frac{1}{2})\lambda^{-\frac{1}{2}} \,,
	\end{split}\end{align}
	with $ c_0=1$ as given in \cite{Watson}. 
	Substituting the arguments of the hypergeometric function that we have in (\ref{finachargr}), we obtain
	the following leading  contributions 
	\begin{eqnarray} \label{asymhyp}
		\lim_{h\rightarrow \infty} {}_2F_1(2-h,1+h,2,\frac{1}{2}-\frac{1}{2\sqrt{1-4Q}}) &\sim& \frac{ e^{h\arcsech(\sqrt{1-4Q})}}{h^{3/2}} \,, \\
		\nonumber
		\lim_{h\rightarrow \infty}  {}_2F_1(3-h,1+h,2,\frac{1}{2}-\frac{1}{2\sqrt{1-4Q}}) &\sim&
		\frac{ e^{h\arcsech(\sqrt{1-4Q})}}{h^{3/2}} \,.
		\end{eqnarray}
		
		Finally we can substitute the asymptotic form in (\ref{asymhyp}) and take the large $h$ limit in 
		the rest of the terms in the thermal one point function given in (\ref{finachargr}) which results in 
		\begin{equation}
	\lim_{h\rightarrow \infty - i \epsilon}
	{\cal O} \rangle\sim e^{-h\log\big[\frac{4}{\sqrt{1-4Q}}\big]+h\arcsech(\sqrt{1-4Q})}e^{-i\pi h}\big[1+e^{-i2\pi h}+\cdots\big]\times(\text{\small powers of h}) .
	\end{equation}
	Here we have followed the prescription in \cite{Grinberg:2020fdj} and 
	 given a small imaginary part to $h$ so as to pick up the phase $e^{-i\pi h}$. 
	Now replacing $h$ by the mass  in the large $h$ or mass limit we obtain 
	\begin{align} \label{finalgeomch}
	\langle {\cal O} \rangle\sim \exp\bigg[m\bigg({{-\frac{R}{d}\log\bigg[\frac{4}{\sqrt{1-4Q}}\bigg]}+{\frac{R}{d}\arcsech(\sqrt{1-4Q})}}-i{\pi \frac{R}{d}}\bigg)\bigg] \, .
	\end{align}

	\subsection*{Comparison with geometric lengths}
	
	Let us now evaluate the lengths $\tau_s, \ell_{\rm hor}, \ell_{\rm sing}$. 
	The proper time between the outer and inner horizons is given by 
	\begin{equation}
	\tau_s = \int_{z_{\rm in }}^{z_{\rm out}}  \frac{R dz }{z \sqrt{ \frac{z^d}{z_0^d} - q z^{2d -2}  - 1}} \,.
	\end{equation}
	We  re-write this integral in terms of $w$  defined in (\ref{defw}) and 
	take the large $d$ limit  of (\ref{dinfty}) 
	to obtain
	\begin{eqnarray}
	\tau_s &=& \frac{R}{d \sqrt{Q}}  \int_{w_{\rm out}}^{w_{\rm in}}  \frac{dw}{w\sqrt{ ( w- w_{\rm out} ) ( w_{\rm in} - w) }} \,, 
	\\ \nonumber
	&=& \frac{\pi R}{d} \,.
	\end{eqnarray}
	The  proper length from the outer horizon to infinity is given by 
	\begin{eqnarray}
	\hat \ell_{\rm hor} &=& \lim_{\epsilon\rightarrow 0}
	\frac{R}{d \sqrt{Q} } \int_{\big( \frac{\epsilon}{z_0} \big)^d }^{w_{\rm out}}  \frac{dw}{ w\sqrt{ ( w_{\rm out}  - w)(w_{\rm in} - w)} } \,, \\ \nonumber
	&=& \frac{R}{d} \log\left( \frac{4}{ \sqrt{ 1- 4Q} } \right)  - R \log \big( \frac{\epsilon}{z_0} \big) \,.
	\end{eqnarray}
	Again the regularised length is obtained by ignoring the divergence, note that from (\ref{lhor}), we see that 
	the divergence is 
	identical to that for the planar  Schwarzschild black hole.
	\begin{equation}
	\ell_{\rm hor} =  \frac{R}{d} \log\left( \frac{4}{ \sqrt{ 1- 4Q} } \right) \,.
	\end{equation}
	The proper length from the inner horizon to the singularity is given by 
	\begin{eqnarray}
	\ell_{\rm sing} &=& 
	\frac{R}{d \sqrt{Q}} \int_{w_{\rm in} }^{\infty}  \frac{dw}{ w\sqrt{ ( w_{\rm out}  - w)(w_{\rm in} - w)} } \,,
	\\ \nonumber
	&=& \frac{R}{d} \arcsech(\sqrt{ 1- 4Q} ) \,.
	\end{eqnarray}
	We use these geometric lengths to rewrite  the one point function obtained in (\ref{finalgeomch})  in the 
	following form 
		\begin{equation}
			\langle {\cal O} \rangle  \sim \exp ( - m ( \ell_{\rm hor} + \ell_{\rm sing } - i \tau_{\rm s }) ) \,.
	\end{equation}
	This is the generic structure of the one point function which was argued using 
	a saddle point approach in \cite{Grinberg:2020fdj}.  This involved a choice of contour in the complex radial 
	coordinate. 
	Here we have obtained it by explicitly evaluating the one point function which involves an 
	integral from the outer horizon to infinity exactly at large $d$.
	  The one point function contains the information of the inner horizon. 
	In the appendix \ref{appendixc}, we show that the same structure of the one point function is obtained for the
	Weyl tensor squared coupling.

\section{Hyperbolic  black holes} \label{hyperbhsec}

In this section, we  consider the  hyperbolic black hole in $AdS_{d+1}$ 
\cite{Emparan:1998he,Birmingham:1998nr} whose metric 
is given by 
\begin{align} \label{hyperbhmetric}
ds^2=\frac{R^2}{z^2}\bigg[-\bigg(1-\frac{z^2}{R^2}\bigg)dt^2+\frac{dz^2}{1-\frac{z^2}{R^2}}+ R^2(du^2+\sinh^2u\ d\Omega_{d-2}^2)\bigg] ,
\end{align}
where $R$ is the radius of $AdS_{d+1}$ and $\Omega_{d-2}$ refers to the sphere $S^{d-2}$. 
The coordinate $z $ is related to the usual radial co-ordinate by $z = \frac{R^2}{r}$. 
This metric can be obtained from a hyperbolic slicing of pure $AdS_{d+1}$
and can be interpreted as a black hole with a horizon 
at $z = R$  with Hawking temperature
\begin{equation}
T = \frac{1}{2\pi R} \,.
\end{equation}
This space is holographically dual to a  conformal field theory on 
$R\times AdS_{d} $ \cite{Emparan:1999gf,Casini:2011kv}. 
We would like to repeat the analysis done in the previous sections for this background and check 
if indeed the one point function contains information of the time from the horizon to singularity. 

We consider the effective action in (\ref{gbaction}) with Gauss-Bonnet coupling. 
The Gauss-Bonnet coupling in this case is a constant which is given by 
\begin{align}\label{gbhyper}
\mathcal{L}_{GB}=\frac{c}{R^4}\,, \qquad\qquad c = (d-2) (d-1) d (d+1) \,.
\end{align}

To construct the bulk-boundary propagator, we follow the procedure discussed in section .
Since the spatial geometry of the boundary is Euclidean $AdS_{d-1}$, we need to expand the 
bulk-bulk Green's function in terms of  normalizable eigen functions  on $AdS_{d-1}$. 
We consider the expansion 
\begin{equation} \label{greenhyexp}
G(z, z', t, t' u, u', \Omega, \Omega' ) =  \int d \omega d\lambda \sum_{l, \sigma}
 \hat G(z, z', \lambda, l, \sigma)  \phi_{\lambda l\sigma} ( u, \Omega)  \phi^{*}_{\lambda l\sigma} ( u', \Omega') 
  e^{-i\omega( t-t') } \,,
\end{equation}
where $\phi_{\lambda l\sigma} $ are eigen functions of the Laplacian on Euclidean $AdS_{d-1}$.  They satisfy
the equation
\begin{equation}
\Box \phi_{\lambda l\sigma} =- \Big(  \lambda^2 + \big( \frac{d-2}{2}  \big)^2 \Big)   \phi_{\lambda l\sigma} \,,
\end{equation}
and the quantum number $\lambda$ is a continuous variable that runs from $0$ to $\infty$. 
The numbers $(l, \sigma)$ refer to quantum numbers on $\Omega_{d-2}$ and $l$ in an integer that 
runs from $0$ to $\infty$, $\sigma$ refer to the other quantum numbers of the spherical harmonics. 
These eigen functions have been explicitly constructed in \cite{Camporesi:1994ga} and they are given by 
\begin{eqnarray}
\phi_{\lambda l \sigma} &=& q_{\lambda, l} ( u) Y_{l\sigma}( \Omega) , \\ \nonumber
q_{\lambda, l} (u) &=& ( i \sinh u)^l {}_2F_1\left( i \lambda + \frac{d-2}{2} +l, 
-i \lambda + \frac{d-2}{2} +l, l + \frac{d-1}{2}; - \sinh^2\frac{u}{2} \right) \,.
\end{eqnarray}
What is important for us is their orthonormality property on $AdS_{d-1}$
\begin{equation}
\int du d\Omega_{d-2} ( \sinh u )^{d-2} \; \phi_{\lambda l \sigma} ( u, \Omega) 
 \phi_{\lambda' l' \sigma'} ( u, \Omega) = \delta_{l, l'} \delta_{\sigma,\sigma'} \delta ( \lambda-\lambda') \,,
\end{equation}
where the integral is on $AdS_{d-1}$ with unit radius.
Substituting the expansion of the Green's function given in (\ref{greenhyexp}) and using the metric
(\ref{hyperbhmetric})  we see that the
 coefficients obey the 
differential equation
\begin{eqnarray} \label{hypgdif}
&& \partial_z \left(  \Big( \frac{ R}{z}\Big)^{d-1} \Big( 1- \frac{z^2}{R^2} \Big) \partial_z \hat G(z, z', \omega, \lambda, l, \sigma) \right)  +
\Big( \frac{ R}{z}\Big)^{d-1} \frac{\omega^2}{ 1- \frac{z^2}{R^2} }  \hat G(z, z', \omega, \lambda, l, \sigma) 
 \nonumber  \\
&& - \Big( \frac{ R}{z}\Big)^{d-1} \left[ m^2 \Big( \frac{ R}{z}\Big)^{2}
+ \frac{1}{R^2} \Big(  \lambda^2 + \big( \frac{d-2}{2}  \big)^2  \Big)  \right] 
 \hat G(z, z', \omega, \lambda, l, \sigma) = \frac{1}{R^{d-1}} \delta( z-z') .
\end{eqnarray}
Similarly the bulk-boundary Green's function admits an eigenfunction expansion given by 
\begin{equation}
K( z', t, t' u, u', \Omega, \Omega' ) =  \int d\omega  d\lambda \sum_{l, \sigma}
 \hat K( z',  \omega, \lambda, l, \sigma)  \phi_{\lambda l\sigma} ( u, \Omega)  \phi^{*}_{\lambda l\sigma} ( u', \Omega') 
 e^{-i\omega( t-t') } .
\end{equation}
From (\ref{bbp}) and the eigen function expansions we see that 
\begin{equation} \label{bbphype}
\hat K ( z',  \omega, \lambda, l, \sigma) = \lim_{z\rightarrow 0 } \frac{2\nu}{ z^\Delta}  \hat G(z, z', \omega, \lambda, l, \sigma) \,.
\end{equation}
Then substituting the eigenmode expansion in the expression for  thermal one point function given in(\ref{1ptgen}), we can obtain 
the expectation value of each mode
\begin{eqnarray}
\langle {\cal O}  \rangle_{w, \lambda,  l, \sigma}
= \alpha \int dz' dt' du' d\Omega'  \sqrt{g'}
\hat{K}( z', \omega, \lambda, l, \sigma) e^{i \omega t'} \phi^{*}_{\lambda l \sigma} ( u', \Omega')  {\cal L}_{GB} (z', t', u', \Omega') \,, \nonumber \\
\end{eqnarray}
where the metric is that given in (\ref{hyperbhmetric}). 
From (\ref{gbhyper}), we see that the Gauss-Bonnet term is independent of time or coordinates on $AdS_{d-1}$. 
This implies that we can perform the $t$ integral, furthermore since the bulk-boundary propagator 
is independent of the quantum numbers $l, \sigma$ due to spherical symmetry we can perform the integral 
on the sphere $S^{d-2}$. 
This leads to 
\begin{eqnarray} \label{modeexph}
\langle {\cal O}  \rangle_{w, \lambda,  l, \sigma} = \alpha \,
2\pi \delta ( w) \delta_{l ,0} \delta_{\sigma,0}
\int dz \frac{R^{2d}} {z^{d+1 } }\hat{K}(z,  0, \lambda, 0, 0) {\cal L}_{GB}(z)
 \int du (\sinh u )^{d-2}   q^*_{\lambda,  0} ( u ) \,.  \nonumber \\
\end{eqnarray}
The integral over $u$ needs to be regulated by placing a cut off in the coordinate $u $ which is along the field theory 
directions.  This will yield a function which just depends on the mode  $\lambda$ it does not contain the 
information of the operator ${\cal O} $. 
Therefore, let us define the thermal one point function of the mode $\lambda$, by scaling out the integral over
$u$. 
\begin{eqnarray} \label{resexp}
\hat {\langle {\cal O}  \rangle}_{\lambda} = \alpha 
\int dz \frac{R^{2d} }{z^{d+1 } } \hat K(z,  \lambda) {\cal L}_{GB}(z), \\ \nonumber
\hat K(z,  \lambda)  = \hat K  ( z, 0, \lambda, 0, 0) \,.
\end{eqnarray}
Here it is understood we are in the zero mode sector $\omega=0, l=0, \sigma =0$. 
It is easy to see that in case the rescaled expectation value $\langle \hat {\cal O}  \rangle_{\lambda}$ is independent of 
$\lambda$, then it implies 
that the thermal one point function $\langle O( t, u, \Omega) \rangle  $ is uniform in the field theory directions. 
This is because if 
$\langle {\cal O} ( t, u, \Omega) $  is constant on  $R\times AdS_{d-1}$ , then  each mode expansion of the expectation 
value is given by  
\begin{eqnarray} \label{uniexp}
 \langle {\cal O}  \rangle_{w, \lambda,  l, \sigma}  &=&  
 \int dt du d\Omega_{d-2} 
 ( \sinh u )^{d-2}  \langle {\cal O} ( t , u , \Omega) \rangle  e^{i\omega t}  q^*_{\lambda, l } (u) Y^*_{l \sigma} ( \Omega ) \,, 
 \\ \nonumber
 &=& \langle {\cal O} \rangle 2\pi \delta(\omega)  \delta_{l, 0} \delta_{\sigma, 0}
   R^{d-1}  \int du ( \sinh u )^{d-2} q^*_{\lambda, 0} \,.
 \end{eqnarray}
 In the second line we have used the fact that $\langle {\cal O} ( t, u, \Omega) \rangle  $  is constant on 
 $R\times AdS_{d-1}$.  
 Comparison of (\ref{modeexph}), (\ref{resexp})  and the second line of (\ref{uniexp}) shows that in case 
 $\hat {\langle {\cal O}  \rangle}_{\lambda} $ is independent of $\lambda$, then it can be identified with the 
 uniform expectation value of the operator in the field theory. 
 Therefore we will evaluate the thermal expectation value given in (\ref{resexp}) and show that in the 
 regime of our interest, that is large operator dimensions $\Delta$, the expectation value is independent 
 of $\lambda$ and  it coincides with the 
 uniform expectation value on the boundary field theory on the hyperbolic space. 
 
 To obtain the Green's function $\hat G(z, z', 0, \lambda , 0, 0)\equiv \hat G(z, z', \lambda) $, it is convenient 
 to go to coordinates
 \begin{equation}
 w = \frac{z^2}{R^2} \,.
 \end{equation}
 Then the equation (\ref{hypgdif}) is written as 
 \begin{eqnarray} \label{hypgreen}
 && 4\partial_w \Big[ w^{ (1- \frac{d}{2} ) }  ( 1- w) \partial_w \hat G(w, w', \lambda) \Big]  \\ \nonumber
 &&-  w^{ -(1 +\frac{d}{2} )} \left[  m^2 R^2 + w\Big(  \lambda^2 + ( \frac{d-2}{2})^2 \Big) \right] \hat G(w, w', \lambda)
 = \frac{2}{R^{d-2}} \delta( w-w') \,.
 \end{eqnarray}
 The Green's function is constructed by using the solution to the homogenous equation 
 The solution which is regular at the boundary $z=0$ is given by 
 \begin{align}
\varphi_{\rm inf }(w)=w^{\Delta /2} \, _2F_1\left(-\frac{d}{4}+\frac{\Delta }{2}-\frac{i \lambda }{2}+\frac{1}{2},-\frac{d}{4}+\frac{\Delta }{2}+\frac{i \lambda }{2}+\frac{1}{2};-\frac{d}{2}+\Delta +1;w\right) ,
\end{align}
 while the solution regular at the horizon is given by 
 \begin{align}
\varphi_{\rm hor} (w)=w^{\Delta /2} \, _2F_1\left(-\frac{d}{4}+\frac{\Delta }{2}-\frac{i \lambda }{2}+\frac{1}{2},-\frac{d}{4}+\frac{\Delta }{2}+\frac{i \lambda }{2}+\frac{1}{2};1;1-w\right) .
\end{align}
Then the Greens function which solves (\ref{hypgreen})
 is obtained by the same methods discussed in the appendix \ref{appendixa}. 
 This results in 
 \begin{eqnarray}
\hat G(w,w', \lambda ) &=&A
\times \bigg(\varphi_{\rm inf} (w)\varphi_{\rm hor} (w')\theta(w'-w)+\varphi_{\rm hor} (w)\varphi_{\inf} (w')\theta(w-w')\bigg),
\\ \nonumber
A&=&-\frac{R^{2-d} \Gamma \left(\frac{d}{2}-\Delta \right) \Gamma \left(\frac{1}{4} (-d+2 \Delta -2 i \lambda +2)\right) \Gamma \left(\frac{1}{4} (-d+2 \Delta +2 i \lambda +2)\right)}{2 \Gamma \left(\frac{1}{2} (d-2 \Delta )\right) \Gamma \left(\frac{1}{2} (2 \Delta -d)+1\right)} \,.
\end{eqnarray}
The bulk-boundary Greens' function can be read out using (\ref{bbphype})
\begin{equation}
\hat K( w, \lambda) = \frac{2\nu}{R^\Delta} A \varphi_{\rm hor} ( w) \,.
\end{equation}
Substituting this in the expression for the thermal one point function in (\ref{resexp}), we obtain 
\begin{equation}
\hat {\langle {\cal O}  \rangle}_{\lambda} = \alpha  \frac{2\nu}{R^\Delta} \frac{c}{2 R^4} AR^d \int _0^1 dw w^{ - (\frac{d}{2} + 1)}
\varphi_{\rm hor} (w) \,.
\end{equation}
We change variables from $w$ to $1-w$ to re-write the integral as
\begin{eqnarray}
\hat {\langle {\cal O}  \rangle}_{\lambda} &=&  \alpha  \frac{2\nu}{R^\Delta} \frac{c}{2 R^4} AR^d \times  \\ \nonumber
&& \quad \int_0^1 dw\ (1-w)^{-\frac{\Delta }{2}-1}  \, _2F_1\left(\frac{1}{4} (d-2 \Delta -2 i \lambda +2),\frac{1}{4} (d-2 \Delta +2 i \lambda +2);1;w\right) ,
\end{eqnarray}
We can perform the integral by using the formula
\begin{eqnarray}
\int_0^1  (1-w)^a \, _2F_1(b,c;1;w) \, dw &=&\frac{\Gamma (a+1) \Gamma (a-b-c+2)}{\Gamma (a-b+2) \Gamma (a-c+2)} \,,
\\ \nonumber
&& \text{ if } {\rm Re}\, (a)>-1 \ \text{and}\,  {\rm Re}\, (a)+2> {\rm Re}\, (b+c)\,,
\end{eqnarray}
which results in 
\begin{eqnarray}
 \hat {\langle {\cal O}  \rangle}_{\lambda} & =&  - \alpha  \frac{2\nu}{R^\Delta} \frac{c}{ R^2} \times 
 \\ \nonumber 
 && \frac{ \Gamma \left(-\frac{\Delta }{2}\right) \Gamma \left(\frac{\Delta -d}{2}\right) \Gamma \left(\frac{1}{4} (-d+2 \Delta -2 i \lambda +2)\right) \Gamma \left(\frac{1}{4} (-d+2 \Delta +2 i \lambda +2)\right)}{4 \Gamma \left(-\frac{d}{2}+\Delta +1\right) \Gamma \left(\frac{1}{4} (-d-2 i \lambda +2)\right) \Gamma \left(\frac{1}{4} (-d+2 i \lambda +2)\right)}\,.
 \end{eqnarray}
 Observe the factor $R^{\Delta +2}$ provides the right scaling dimension for the expectation value. 
 
 Now that we have the exact expectation value of the operator, we can take the large $\Delta \rightarrow \infty -i\epsilon$
 limit with $\lambda$ fixed  we obtain
 \begin{equation}
 \lim_{\Delta \rightarrow \infty -i\epsilon}   \hat {\langle {\cal O}  \rangle}_{\lambda} \sim 
  \csc \bigg(\frac{\pi  \Delta }{2}\bigg)\ e^{-\Delta  \log (2)} \times (\text{powers of }\Delta) \,.
 \end{equation}
 From our earlier discussion, we can conclude that 
 since this value  is independent of $\lambda$, we can conclude that in the large $\Delta$ limit the operator 
 acquires uniform expectation value on the field theory directions  $R\times AdS_{d-1}$. 
 Using $\Delta \sim m R$ in the large $\Delta$ limit and ignoring corrections due to powers of $\Delta$, 
  we can re-write this uniform thermal expectation value as
 \begin{equation}\label{finalhypero}
  \lim_{\Delta \rightarrow \infty -i\epsilon}   \langle {\cal O} \rangle \sim 
  \exp \Big[ m \big(  - \frac{i \pi R }{2}  - R \log 2  \big)  \Big] .
 \end{equation}

\subsection*{Comparison with geometric lengths}

At this stage we can compare the form of the thermal expectation value in (\ref{finalhypero}) with the 
expression proposed in \cite{Grinberg:2020fdj} in terms of geometric lengths. 
The proper time from the horizon to singularity for the hyperbolic black hole in (\ref{hyperbhmetric}) is given by 
\begin{equation}
\tau_s= R\int_R^\infty \frac{ dz}{ z \sqrt{ \frac{z^2}{R^2 } -1}}  =\frac{\pi  R}{2}\,.
\end{equation}
Recall that the radial co-ordinate $r = \frac{R^2}{z}$. 
Note that unlike the planar Schwarzschild black hole, this length is independent of the dimension $d$. 
The proper length from infinity to the horizon is given by 
\begin{eqnarray}
\hat \ell_{\rm hor} &=& \lim_{\epsilon\rightarrow 0} R \int_\epsilon^R \frac{dz}{ z \sqrt{ 1 - \frac{z^2}{R^2} }} \,, \\ \nonumber
&=& R \log(2) - R\log \Big( \frac{\epsilon}{R} \Big) \,.
\end{eqnarray}
The regularized proper length is obtained by ignoring the divergence, so we obtain
\begin{equation}
\ell_{\rm hor} = R \log(2) \,.
\end{equation}
Using the proper time  and the regularized proper length, we can write the expectation value in 
(\ref{finalhypero}) as 
\begin{equation}
 \lim_{m \rightarrow \infty -i\epsilon}   \langle {\cal O} \rangle \sim \exp ( -im \tau_s - m \ell_{\rm hor} ) \,.
 \end{equation}
 Therefore we see that the 
geometric form for the expectation value for operators of large dimensions is true even in the presence of 
black holes with hyperbolic horizons. 
One of the consequences of our calculation is that generically, we expect operators of large dimensions on 
conformal field theories on spatial hyperbolic surfaces to obtain expectation values at finite temperature.

\section{BTZ black holes with angular momentum}
\label{section4}

\begin{figure}
		\centering
	\begin{tikzpicture}[scale=1.2]
	\draw[color=blue,fill=black] (0,0) circle (.6cm);
	\draw (0,0) circle (1.6cm);
	\draw[color=blue] (0,0) circle (.8cm);
	\draw[color=red] (.8,0) -- (1.6,0)  node[right] {$ \varphi $};
	\node[color=blue] (c) at (.75,.75){$ \chi $};
\end{tikzpicture} 	
\caption{One point function induced by the cubic coupling $\chi^2 \varphi$ in the BTZ background \cite{Kraus:2016nwo}.}
\label{figbtz}
\end{figure}
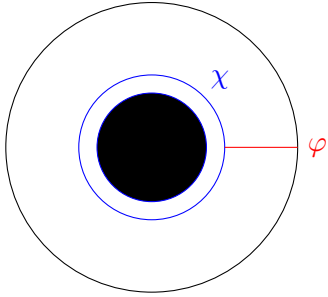

In \cite{Grinberg:2020fdj},  the examples considered did not include the rotating black holes, moreover it is not clear that 
the saddle point arguments  to arrive at the thermal expectation value 
provided in \cite{Grinberg:2020fdj} for black holes with inner horizons generalise to that of black holes
with rotation. This is because these metrics   which was used in the 
saddle point arguments had spherical symmetry.
In this section we use the  rotating BTZ  black hole geometry to study the behaviour of thermal one point function
of scalars dual to primaries in the CFT.  Though many quantities are exactly solvable in this geometry, 
one apparent difficulty with this is that  in the dual field theory
  conformal invariance  in $2d$ ensures that primary operators do not acquire non-trivial expectation 
values at finite temperature if the CFT is on a real line. 
In the  3 dimensional dual holographic description, this is reflected in the fact that 
both the Gauss-Bonnet curvature  as well  Weyl tensor curvature square vanishes in $3d$.
Therefore the mechanism discussed in the previous sections does not apply to holographic  $2d$ CFTs. 
However it is known  that one point functions of CFTs  on a torus  do acquire non-trivial expectation value. 
In \cite{Kraus:2016nwo}, this expectation value was interpreted holographically as arising due to the  cubic coupling 
of the field $\varphi$  dual to the operator of interest with another bulk  field $\chi$. 
More explicitly one considers the  following action 
\begin{align}\label{btzact}
S=\frac{1}{16\pi G_N}\int d^{3}x \sqrt{g}\ (\partial_\nu\varphi\partial^\nu\phi+m^2\varphi^2+\partial_\nu\chi\partial^\nu\chi+\mu^2\chi^2+g \chi^2\varphi) \,.
\end{align}
The field $\chi^2$ acquires non-trivial expectation value due to the sum of propagators that wind around the 
horizon of the BTZ black hole, this in turn induces a one point function for the field $\varphi$. 
This is shown in figure  \ref{figbtz}. 

In \cite{Grinberg:2020fdj}, the action  in (\ref{btzact}) was used to 
 show that  for the Schwarzschild BTZ black hole, the leading contribution to the expectation value of the 
operator ${\cal O} $ dual to $\varphi$  is given by 
\begin{eqnarray}\label{btzexpt}
\lim_{\stackrel{m \rightarrow \infty -i \epsilon}{ L\rightarrow \infty} }
\langle {\cal O} \rangle  &\sim  &
( {\rm powers\; of} \;m)  \exp ( - im \tau_s - m \ell_{\rm hor} )   \times e^{  - \frac{ 2\pi L}{\beta} } 
\\ \nonumber
&& 
+ 
O (e^{- \frac{4\pi L}{\beta} } ) \,.
\end{eqnarray}
Here $L$ is the periodicity in the spatial direction,  when $L\rightarrow \infty$, the expectation value 
vanishes.
In this section we will show that when one considers the BTZ black hole with rotation, the 
thermal one point function behaves as given in (\ref{btzexpt}), where now 
$\tau_s, \ell_{\rm hor}$ are evaluated in the rotating BTZ geometry.
 Note that we do not see the occurrence 
of ${\rm \ell_{sing}}$, the proper length from the inner horizon to the origin, 
 unlike what was seen for the black hole with charge.  
 
To see if this behaviour holds in other situations, we evaluate the expectation value of the composite made
out of the bilinear ${\cal O} ^2$ in 2d CFT on a line but held at distinct left and right temperatures. 
The dual description of this CFT at large central charge is the BTZ black hole with angular momentum. 
Since the  field is a composite, its expectation value at finite temperature is non-vanishing and can be obtained 
using conformal invariance. 
A closed expression can be obtained for this expectation value when the operator $O$ has large integer dimensions. 
We show that this expectation value can be written as 
\begin{equation}
\lim _{m \rightarrow \infty - i \epsilon} \langle {\cal O} ^2\rangle \sim  \exp (-2  i m \tau_s - 2m \ell_{\rm hor} ) \,.
\end{equation}
Again we do not see the appearance of the proper length $\ell_{\rm sing}$. 
It will be interesting to see if this is a feature true for thermal one point functions in rotating black holes in 
higher dimensions. 

\subsection{Thermal one point function from gravity}

We consider the BTZ metric with angular momentum which is given by 
\begin{align}\label{metricbtz}
ds^2=\frac{\left(-r^2+r_-^2+r_+^2\right) dt^2}{R^2}+\frac{r^2 R^2 dr^2}{\left(r^2-r_-^2\right) \left(r^2-r_+^2\right)}+r^2 d\phi ^2-\frac{2 r_- r_+ dt d\phi }{R} \,.
\end{align}
Then by AdS/CFT, the expectation value of the operator dual to the scalar $\varphi$ is given by 
\begin{equation} \label{1ptbtz}
\langle {\cal O} (t, \phi ) \rangle = g \int \sqrt{g} d^3 x' \langle \chi^2 ( x') \rangle K( x', t, \phi ) \,.
\end{equation}
where $K(x', t, \phi)$ is the bulk-boundary propagator of the $\varphi $ field  in the $BTZ$ background, and the integral is 
carried out in the BTZ geometry outside the horizon. 
The expectation value $\langle \chi^2(x') \rangle $ is obtained by 
considering the bulk-bulk propagator of the $\chi$ field and subtracting the coincident limit singularity. 
We now provide the  details for  evaluation of the expectation value $\chi^2$, the  bulk-boundary propagator  $K(x', t, \phi)$  and then 
proceed to evaluate the integral in (\ref{1ptbtz}). 

\subsubsection*{Bulk-Bulk propagator for $\chi$}

Bulk-Bulk propagators between 2 points 
 in global $AdS_3$ are known exactly in terms of the geodesic  distance between the points. 
 Since the $BTZ$ is a quotient of global $AdS_3$ through the identification $\phi \rightarrow \phi + 2\pi$, 
 we first write down the geodesic length between the following two points at the same radial position and time 
 coordinate in BTZ  but at angular separation of $\rho$.  
 \begin{equation} \label{2points}
 x_1:   ( t, r, \phi =0) \, , \qquad x_2 : ( t, r, \phi = \rho) \,.
 \end{equation}
 The easiest approach to find the geodesic length is to use the embedding of the BTZ background into 
 the hyperboloid defined by 
 \begin{equation}
 -T_1^2-T_2^2+X_1^2+X_2^2=-R_{}^2 .
 \end{equation}
 Then the BTZ black hole  with angular momentum in (\ref{metricbtz}) is obtained  from the metric 
 \begin{align}
ds^2=-dT_1^2-dT_2^2+dX_1^2+dX_2^2 \,.
\end{align}
with the following embedding  \cite{Banados:1992gq}. 
\begin{align}
T_1=\sqrt{\frac{r^2-r_+^2}{r_+^2-r_-^2}} R \sinh \left(\frac{r_+ t-r_- R \phi }{R^2}\right) ,\\ \nonumber
T_2=\sqrt{\frac{r^2-r_-^2}{r_+^2-r_-^2}} R \cosh \left(\frac{r_+ R \phi -r_- t}{R^2}\right) ,\\ \nonumber
X_1=\sqrt{\frac{r^2-r_+^2}{r_+^2-r_-^2}} R \cosh \left(\frac{r_+ t-r_- R \phi }{R^2}\right) ,\\ \nonumber
X_2=\sqrt{\frac{r^2-r_-^2}{r_+^2-r_-^2}} R \sinh \left(\frac{r_+ R \phi -r_- t}{R^2}\right) .
\end{align}
From this relations, we see that the location of the  2 points in (\ref{2points}),  in terms of the co-ordinates
$( T_1 (x_i)), T_2 (x_i) , X_1 (x_i) , X_2(x_i) )$ can be obtained. 
Then the geodesic distance $\hat d$  between these points is obtained  using the relation
\begin{equation}
-R^2 \cosh \hat d =  -T_1( x_1) T_2 ( x_2)  - T_2(x_1) T_2( x_2) + X_1( x_1) X_1( x_2)  + X_2( x_1) X_2( x_2) \,.
\end{equation}
We obtain the following  simple formula for the geodesic distance between the points (\ref{2points}) 
\begin{equation}
\cosh \hat{d}=\frac{\left(r^2-r_-^2\right) \cosh \left(\frac{\rho  r_+}{R}\right)+\left(r_+^2-r^2\right) \cosh \left(\frac{\rho  r_-}{R}\right)}{r_+^2-r_-^2} \,.
\end{equation}

The bulk-bulk propagator in $AdS_3$ is given by  \cite{DHoker:1999mqo}
\begin{align}
G_\Delta (u) =\frac{C_\Delta2^\Delta}{u^\Delta} \ _2F_1\left(\Delta,\Delta-\frac{1}{2},2\Delta-1,-\frac{2}{u}\right) ,
\end{align}
where 
\begin{eqnarray} \label{defu}
&&u=2 \sinh^2\frac{\hat{d}}{2}  = \frac{(r_+^2-r^2) \cosh \left(\frac{ \rho  r_-}{R} \right)+(r^2-r_-^2)
 \cosh \left(\frac{\rho  r_+}{R} \right)-(r_+^2-r_-^2)}{r_+^2-r_-^2} \,, \\ \nonumber
 &&\qquad c_{\Delta} =  
\frac{\Gamma(\Delta) \Gamma( \Delta - \frac{1}{2} ) }{ ( 4\pi)^{\frac{3}{2}}  \Gamma( 2 \Delta -1) } \,.
\end{eqnarray}
We will consider the case when the field $\chi$ has conformal dimensions $\Delta =1$, then 
\begin{equation}
G_{1}(r, \rho)  = \frac{1}{8\pi} \frac{1}{ \sqrt{ \frac{u}{2} ( \frac{u}{2} +1) } } \,.
\end{equation}
Substituting for $u$ from (\ref{defu}) we obtain 
\begin{equation} \label{expchi1}
G_1(r,  \rho)  = \frac{1}{4\pi} \frac{(r_+^2-r_-^2)}{\sqrt{\big[(r_+^2-r^2)\cosh(\frac{ r_- \rho}{R} )+(r^2-r_-^2)\cosh(\frac{ r_+\rho}{R} )\big]^2-(r_+^2-r_-^2)^2}} \,.
\end{equation}
Using this bulk to bulk propagator, we obtain 
\begin{equation} \label{expchi2}
\langle \chi^2( r )\rangle = \sum_{n = -\infty, n\neq 0 }^\infty G_1( 2\pi n)  = 2 \sum_{n=1}^\infty G_1( 2\pi n ) \,.
\end{equation}
Here we have removed the coincident singular limit and summed over all the windings to ensure periodicity 
in $\phi \rightarrow \phi +2\pi$. 
Note that the expectation value of $\chi^2$ just depends on the radial position. 

\subsubsection*{The bulk-boundary propagator for $\varphi$}

From the fact that the expectation value of $\chi^2$ depends only the radial position, it is convenient 
to work in the Fourier expansion of the bulk-boundary propagator in frequency and the angle $\phi$. 
Let 
\begin{equation}
K( r', t, t', \phi,\phi')
  = \sum_{n=-\infty}^{\infty } \int \frac{d \omega }{2\pi} e^{ - i \omega ( t-t') + i n ( \phi -\phi') } \hat K ( r, r', \omega, n) \,.
\end{equation}
Then substituting this Fourier expansion in (\ref{1ptbtz}) and using the fact that $\langle \chi^2 (r)\rangle  $ just 
depends on the radial position, we can perform the integral over $t', \phi'$. 
This leads to
\begin{equation} \label{int1pbtz}
\langle {\cal O} (x, t) \rangle = g  \int_{r=r_+}^\infty  r' dr' \hat K ( r', 0, 0) \langle \chi^2( r' )\rangle \,.
\end{equation}
We obtain the zero mode of the bulk-boundary propagator $\hat K( r', 0, 0)$ by examining the 
zero mode of the bulk-bulk Green's function, which satisfies the differential equation
\begin{equation}
\frac{1}{r} \partial_r \left( \frac{ ( r^2 - r_-^2) ( r^2 - r_+^2) }{ r R^2} \partial_r \hat G( r, r', 0, 0)  \right) 
- m^2  \hat G( r, r', 0, 0)  = \frac{\delta ( r- r') }{r} \,.
\end{equation}
We change variables to 
\begin{equation} \label{defx}
x = \frac{ r^2 - r_+^2}{ r^2 - r_-^2} \,.
\end{equation}
Then the differential equation for the Green's function reduces to 
\begin{equation}
4 ( 1-x)^2 \partial_x \left[  x \partial_x \hat G( x, x') \right]
 - m^2 R^2  \hat G ( x, x') =\frac{ 2 ( 1-x)^2 R^2}{r_+^2 -r_-^2} \delta (x-x') \,,
\end{equation}
here it is understood that $\hat G(x, x')$ refers to the Fourier zero mode along $t, \phi$ directions.

We construct the Green's function by solving the homogenous equation. 
Let us define $ h = \frac{\Delta}{2} $, which is related to the mass by 
\begin{equation} \label{btzhdef}
h = \frac{1}{2} ( 1 + \sqrt{ 1 + m^2 R^2 } ) \,,  \qquad  \nu = \sqrt { 1+ m^2 R^2} \,.
\end{equation}
The solution which is well behaved near the outer horizon $r=r_+, x=0$  is given by 
\begin{equation}
\varphi_{\rm hor} = (1-x)^{1-h} \, _2F_1(1-h,1-h;1;x)\,,
\end{equation}
and the solution which is well behaved at the boundary  $r=\infty, x=1$ is given by 
\begin{equation}
\varphi_{\rm inf} =
(1-x)^h \, _2F_1(h,h;2 h;1-x) \,.
\end{equation}
Then the Green's function is given by 
\begin{eqnarray}
G(x,x') &=&  A \times \Big[ \varphi_{\rm inf} (x) \varphi_{\rm hor}(x')  \theta (x-x') 
+ \varphi_{\rm hor} ( x) \varphi_{\rm inf} ( x') \theta (x'- x) \Big] , \\ \nonumber
A &=&-\frac{R^2 \Gamma (h)^2}{2. \Gamma (2 h). \left(r_+^2-r_-^2\right)} \,.
\end{eqnarray}
We can now use the bulk-bulk Green's function to obtain the 
bulk-boundary Green's  function. 
Substituting for $z = \frac{R^2}{r}$ in (\ref{bbp})  and then using (\ref{defx}) we obtain 
\begin{eqnarray}\label{bbbtz}
K(x) &=& \lim_{x'\rightarrow 1}  \frac{ 2\nu }{R^{4h}} \frac{ ( r_+^2 - r_-^2 )^h}{ ( 1-x') ^h} G(x,x'), \\ \nonumber
&=& \frac{2\nu A}{ R^{4h}}   ( r_+^2 - r_-^2 )^h ( 1- x)^{1-h}  \, _2F_1(1-h,1-h;1;x).
\end{eqnarray}

\subsubsection*{Thermal one point function}

We are  ready to substitute the necessary ingredients in (\ref{int1pbtz}) to obtain the one point function 
Using the bulk-boundary Green's function (\ref{bbbtz}),  the expectation value 
$\langle \chi^2 (r) \rangle$ from  (\ref{expchi1}) 
we obtain 
\begin{eqnarray} \label{1ptbtz2}
\langle {\cal O }\rangle &=& -   {\cal K} (h)
\sum_{n=1}^\infty 
 \frac{ 1 }{ \sinh(\frac{ 2\pi  n r_+}{R}) }
  \int_0^1 dx \frac{ (1-x)^{h-1}\ _2F_1(h, h , 1;x) }{ \sqrt{\big(1-\hat \chi_n^{(1)} x \big)\big(1-\hat \chi_n^{(2)} x\big)} } \,, \\ \nonumber
  {\cal K} (h)& =&   \frac{g \nu R^2}{4\pi } \frac{ (r_+^2-r_-^2)^h  (\Gamma(h) )^2}{ R^{4h} \Gamma( 2h) } \,.
\end{eqnarray}
where we have changed the integration variables to $x$ using (\ref{defx}) and 
\begin{equation}
\hat \chi_n^{(1)}=\frac{\cosh^2(\frac{\pi n r_-}{R})}{\cosh^2(\frac{\pi n r_+}{R})} \,,\ \ \ \text{and} \ \ \ \ 
\hat \chi_n^{(2)}=\frac{\sinh^2(\frac{\pi n  r_-}{R} )} {\sinh^2(\frac{\pi  n r_+}{R})} \,.
\end{equation}
In (\ref{1ptbtz2}),  we have used the fact that the expectation value of the operator is uniform from (\ref{int1pbtz}) and dropped
the dependence on $(t, \phi)$. 
To take various limits it is convenient to change variables to $w = 1-x$ in (\ref{1ptbtz2}) in the integrand,  this results in 
\begin{eqnarray} \label{1ptbtz22}
\langle {\cal O }\rangle &=& -   {\cal K} (h)
\sum_{n=1}^\infty 
 \frac{ 1 }{ \sinh^2 (\frac{ \pi  n r_+}{R}) -  \sinh^2 (\frac{ \pi  n r_-}{R})  }
  \int_0^1 dw \frac{ w^{h-1}\ _2F_1(h, h , 1;1-w) }{ \sqrt{\big(1- \chi_n^{(1)} w \big)\big(1-\chi_n^{(2)} w\big)} } \,,
  \end{eqnarray}
where
\begin{eqnarray}
{\chi}_n^{(1)} &=&\frac{\hat \chi_n^{(1)}  }{ \hat \chi_n^{(1)} -1}=\frac{\cosh ^2\left(\frac{\pi  n r_-}{R}\right) }
{ \sinh^2 (\frac{ \pi  n r_-}{R}) -  \sinh^2 (\frac{ \pi  n r_+}{R}) }, \\ \nonumber
{\chi}_n^{(2)} &=&\frac{\hat \chi_n^{(2) } }{\hat \chi_n^{(2)}-1}=\frac{\sinh ^2\left(\frac{\pi n   r_-}{R}\right)}{
 \sinh^2 (\frac{ \pi  n r_-}{R} ) -  \sinh^2 (\frac{ \pi  n r_+}{R} ) } \,.
\end{eqnarray}
We can now expand the denominator in (\ref{1ptbtz22}) using 
\begin{align}
\frac{1}{\sqrt{(1-{\chi}_1w)(1- {\chi}_2 w)}}
=\sum _{k=0}^{\infty } P_k\left(\frac{{\chi} _1+{\chi} _2}{2 \sqrt{{\chi} _1 {\chi} _2}}\right) \left(\sqrt{ {\chi} _1 {\chi} _2}\  w\right)^k ,
\end{align}
where $P_k$ refers to the Legendre polynomial of order $k$. 
Substituting this expansion in (\ref{1ptbtz2}),  and integrating term by term we obtain 
\begin{eqnarray} \label{1ptbtz3}
\langle {\cal O }\rangle &=&    {\cal K} (h) 
\sum_{n=1,  k=0}^\infty 
 \frac{ 1 }{ \sinh^2 (\frac{ \pi  n r_+}{R}) -  \sinh^2 (\frac{ \pi  n r_-}{R})  } \times  \\ \nonumber
 && \qquad\qquad \left[
  P_k\Big(\frac{{\chi} _n^{(1)} +{\chi} _n^{(2)} }{2 \sqrt{{\chi} _n^{(1)} {\chi} _n^{(2)}}}\Big) \left(\sqrt{ {\chi} _n^{(1)}  {\chi} _n^{(2)} } \ \right)^k
  \frac{\Gamma (-h+k+1) \Gamma (h+k)}{\Gamma (k+1)^2} \right] .
 \end{eqnarray}
 The above result is the closed form expression for the thermal one point function of the 
 operator ${\cal O}$, in the  rotating BTZ background which is induced by the cubic coupling with the field $\chi$.
 It will be interesting to compare this result with that of the CFT by developing the methods in \cite{Kraus:2016nwo}. 
 This work assumed that the black hole could be modelled as a heavy state in the 
 CFT, but the case when the heavy state had different left and right conformal dimensions was not considered.
 
 \subsubsection*{$r_-\rightarrow 0$ limit}
 
In \cite{Grinberg:2020fdj}, the one point function for the Schwarzschild BTZ was obtained.  Let us first reproduce this results from the general 
expression (\ref{1ptbtz3}). 
In the limit $r_-\rightarrow 0$, we see that
  \begin{eqnarray} \nonumber
 &&  {\chi}_n^{(1)} \rightarrow-\frac{1}{\sinh ^2(\frac{\pi n  r_+}{R})}\ \ \ \   \text{and}\ \ \ \ { \chi}_{n}^{(2)} \rightarrow0 \,, \\
  &&P_k\Big(\frac{{\chi} _n^{(1)} +{\chi} _n^{(2)} }{2 \sqrt{ {\chi} _n^{(1)}  {\chi} _n^{(2)} }}\Big) 
  \Big[\sqrt{ {\chi} _n^{(1)}  {\chi} _n^{(2)} }\Big]^k \rightarrow \frac{ \Gamma (2 k+1)}{ 4^k \Gamma (k+1)^2}\ (
  \chi_n^{(1)} )^k, \\ \nonumber
  && {\cal K} ( h) \rightarrow \frac{g\nu R^2}{4\pi} \frac{r_+^{2h} \Gamma(h)^2  }{R^{4h} \Gamma(2h) } \,.
  \end{eqnarray}
  Substituting these leading terms in (\ref{1ptbtz3}), we obtain 
  \begin{eqnarray} \label{1ptbtzsch}
  \langle {\cal O} \rangle&=& \frac{g\nu R^2}{4\pi } \frac{ \Gamma(h)^2  }{z_0^{2h} \Gamma(2h) }  \times \\ \nonumber
  && \qquad
  \sum_{n=1, k=0}^\infty   \frac{1}{\sinh ^2(\frac{\pi n  r_+}{R})}
  \frac{ \Gamma (2 k+1)}{ 4^k \Gamma (k+1)^2}\ (
  \chi_n^{(1)} )^k  \frac{\Gamma (-h+k+1) \Gamma (h+k)}{\Gamma (k+1)^2} \,, \\ \nonumber
  &=& \frac{g\nu R^2}{4\pi} \frac{  \Gamma(h)^2  }{z_0^{2h} \Gamma(2h) }  \sum_{n=1}^\infty
  \frac{\pi  \csc (\pi  h)}{\sinh^2(  \frac{ \pi n r_+ }{R})} \ _3F_2\left(\frac{1}{2},1-h,h;1,1;-\frac{1}{\sinh ^2(\frac{\pi n  r_+}{R})}\right) .
  \end{eqnarray}
  To obtain the second line, we have used the Legendre duplication formula to simplify the Gamma functions in the numerator and 
  then used the definition of the hypergeometric function $\ _3F_2 $ to sum over $k$. 
  We have also replaced $\frac{R^2}{r_+} $ by $z_0$. 
  The last line in (\ref{1ptbtzsch}), is the expression obtained in \cite{Grinberg:2020fdj} for thermal one point function in the 
  Schwarzschild BTZ background. 
  
  To interpret the one point function geometrically, we first take the limit 
 $\frac{ 2\pi r_+}{R}  \rightarrow \infty$. 
 At this point it is instructive to relate this limit to the one taken in  \cite{Grinberg:2020fdj}, in which the length of the identification
 in the spatial direction of the BTZ black hole is large. 
 To do this let us write the metric of the  BTZ Schwarzschild metric from  (\ref{metricbtz})
 \begin{equation}\label{schbtzmet}
 ds^2 =  \frac{R^2}{z^2} \Big[  \big( 1- \frac{z^2}{z_0^2} \big) dt^2 + \frac{dz^2}{ 1- \frac{z^2}{z_0^2} } + R^2 d\varphi^2 \Big] .
 \end{equation}
 Here we have written the Euclidean BTZ, in which $t$ is identified as $t \rightarrow t + \beta$, where 
 \begin{equation} \label{defz0}
 z_0 = \frac{R^2}{r_+} = \frac{\beta}{2\pi} \,,
 \end{equation}
 On  re-defining the co-ordinates as 
 \begin{equation} 
 z' = \frac{z}{z_0} \,, \qquad t' =  \frac{t}{z_0} \,, \qquad x= \frac{R}{z_0} \phi \,.
 \end{equation}
 the metric becomes
 \begin{equation}
 ds^2 = \frac{R^2}{z^{\prime 2}}  \Big[   ( 1- z^{\prime 2} )dt^{\prime 2 } + \frac{ dz^{\prime 2}}{ 1 -  z^{\prime 2}} + dx^2 \Big] .
 \end{equation}
 With this scaling, 
 we have the identifications 
 \begin{equation}
 t'  \sim t' + 2\pi, \qquad x \sim x + 2\pi \frac{2\pi R}{\beta} \,.
 \end{equation}
 To arrive at the identification in $x$ we have used the relation between $z_0$ and $\beta $ in (\ref{defz0}). 
 This is the rescaled co-ordinates used in \cite{Grinberg:2020fdj}.  The periodicity in the spatial direction  can be read 
 from (\ref{schbtzmet}). It is given by 
 to be $L = 2\pi R$ and therefore $x \sim x+ \frac{2\pi L}{\beta}$. 
 Let us now examine the ratio $\frac{r_+}{R}$, from (\ref{defz0}) 
 \begin{equation}
 \frac{2\pi r_+}{R} = 2\pi \frac{ 2\pi R}{\beta} =  \frac{2\pi L}{\beta} \,,
\end{equation}
Therefore taking the limit 
\begin{equation} \label{spacelim}
\frac{2\pi r_+}{R} \rightarrow\infty 
\end{equation}
corresponds to the same limit of large spatial periodicities of \cite{Grinberg:2020fdj}.

Examining the one point function in (\ref{1ptbtzsch}) in the limit (\ref{spacelim}), we see only the $n=1$ term contributes and 
we can also just retain the leading term from the hypergeometric function which results in 
\begin{eqnarray}
 \langle {\cal O} \rangle= \frac{g\nu R^2}{2 } \frac{ \Gamma(h)^2  }{z_0^{2h} \Gamma(2h) } 
 \csc(\pi h) e^{- \frac{2\pi  r_+}{R} } .
\end{eqnarray}
Note that the $z_0^{-2h}$ just provides the scaling dimensions for the expectation value. 
In this form it is easy to see that on further taking the limit $h\rightarrow \infty -i \epsilon$
\begin{eqnarray}
\lim_{h\rightarrow \infty - i \epsilon}
 \langle {\cal O} \rangle &\sim&  e^{- i \pi h} 2^{-2h}  e^{- \frac{2\pi  r_+}{R} },  \\ \nonumber
&  \sim& e^{-i \pi \frac{m R}{2} }  4^{ - \frac{mR}{2}}  e^{- \frac{2\pi  r_+}{R} } .
 \end{eqnarray}
 To obtain the last line we have used (\ref{btzhdef}) to replace $h$ in terms of the mass $m$. 
Comparing the coefficient of $e^{- \frac{2\pi  r_+}{R} }$  in  (\ref{schwexpo}), we see that it 
behaves just as seen for thermal one point functions in higher dimensions.

%
%
%
%
%
%
%
%

\subsubsection*{ Limit $\frac{r_-}{R}, \frac{r_+}{R} \rightarrow \infty$ , with $\frac{r_-}{r_+}$ held fixed}

In this limit, we still can keep track of the dependence of the one point function on the inner horizon 
Before we take this limit, it  is convenient to simplify the dependence on $r_-, r_+$ in (\ref{1ptbtz3}). 
We can write
\begin{equation} \label{simpli1}
\frac{{\chi} _n^{(1)} +{\chi} _n^{(2)} }{2 \sqrt{{\chi} _n^{(1)} {\chi} _n^{(2)}}} = \coth\frac{2\pi n r_-}{R}\,,
\end{equation}
and 
\begin{equation} \label{simpli2}
\sqrt{ {\chi} _n^{(1)}  {\chi} _n^{(2)}  }=  \frac{\sinh \frac{2\pi n r_-}{R }}{ \cosh \frac{2\pi n r_-}{R} - \cosh\frac{2\pi n r_+}{R}} \,.
\end{equation}
We need the limit
\begin{equation} \label{simpli3}
\lim_{\frac{r_-}{R} \rightarrow \infty} P_k( \coth \frac{2\pi n r_-}{R} )  = 1 + O( \exp ( - \frac{4\pi n r_-}{R} ) \,.
\end{equation}
We substitute  (\ref{simpli1}),  (\ref{simpli2}) into the expression for the one point function in (\ref{1ptbtz3}) and take the limit
\begin{equation} \label{btzlim}
\frac{r_-}{R} \rightarrow \infty, \quad \frac{r_+}{R} \rightarrow \infty, \quad \frac{r_-}{r_+} =  \delta \;\; {\rm fixed} .
\end{equation}
Using (\ref{simpli3}), we see that the leading contribution arises from $n=1, k=0$ term in the sum. 
All the remaining terms are exponentially suppressed compared to this term, which is given by 
\begin{eqnarray}
\langle {\cal O} \rangle &= &\frac{g\nu R^2}{2}
\frac{  ( r_+^2 - r_-^2)^h  \Gamma(h)^2}{ R^{4h} \Gamma(2h) } \csc(\pi h) e^{ - \frac{2\pi r_+}{R}}  + 
O\Big[  \exp ( - \frac{2\pi r_+}{R} ( 2- \delta) ) \Big] , \nonumber \\
&=& \frac{g\nu R^2}{2 } \frac{ \Gamma(h)^2  }{z_0^{2h} \Gamma(2h) } \csc(\pi h) 
e^{h \log ( 1- \delta^2) }  e^{ - \frac{2\pi r_+}{R}} ,
\end{eqnarray}
with $ z_0=\frac{R^2}{\sqrt{{r_+}^2-r_-^2}} $.\\\\
Finally we can take the $h\rightarrow \infty -i\epsilon$ 
\begin{eqnarray} \label{btzfinal1pt}
\lim_{h \rightarrow \infty - i \epsilon} \langle {\cal O} \rangle \sim
e^{- i \pi h}   e^{-h ( \log 4   -\log ( 1- \delta^2) }  e^{ - \frac{2\pi r_+}{R}} , \\ \nonumber
= e^{-i  \frac{\pi mR}{2}}  e^{ - \frac{mR}{2} ( \log 4  - \log( 1- \delta^2) )} e^{ - \frac{2\pi r_+}{R}} .
\end{eqnarray}

\subsubsection*{Comparison with geometric lengths}

From the metric (\ref{metricbtz},  the time to singularity is given by the integral 
\begin{eqnarray}\label{btztaus}
\tau_s &=& R \int_{r_-}^{r_+} \frac{ rdr}{\sqrt{ ( r_+^2 - r^2 ) ( r^2 - r_-^2)  }} \,, \\ \nonumber
&=& \frac{R\pi}{2} \,.
\end{eqnarray}
The proper length from infinity to the horizon is given by 
\begin{eqnarray}
\hat \ell_{\rm hor} &=& \lim_{\epsilon\rightarrow 0} R \int_{r_+}^{\frac{R^2}{\epsilon}}  \frac{rdr}{ \sqrt{ ( r_+^2 - r^2 ) ( r^2 - r_-^2)  }} \,
,  \\ \nonumber
&=& -\frac{R}{2} \log( 1- \delta^2)  + R\log(2)  - \log \frac{\epsilon}{z_0} \,, \qquad  \qquad z_0 = \frac{R^2}{r_+} \,.
\end{eqnarray}
Here we have chosen the same cut off as in (\ref{cutoffdef}), since $r$ and $z$ are related by $r = \frac{R^2}{z}$. 
Therefore the regularized proper length is given by
\begin{equation} \label{lhorbtz}
\hat \ell_{\rm hor} = - \frac{R}{2} \log ( 1- \delta^2)  + R \log(2) \,.
\end{equation}
Finally one can also evaluate the proper length from the inner horizon to the singularity, we obtain 
\begin{equation}
\hat \ell_{\rm sing} = \frac{R}{2} \log \Big( \frac{1-\delta}{1+ \delta} \Big) .
\end{equation}

Comparing the leading contribution to the thermal one point function in (\ref{btzfinal1pt}) with these geometric lengths, 
 we see that it can be written as 
\begin{equation}
\langle {\cal O} \rangle \sim \exp ( - i m  \tau_s  - m \ell_{\rm hor})   e^{ - \frac{2\pi r_+}{R}} .
\end{equation}
Note that in the limit (\ref{btzlim}) we do not see the presence of $\ell_{\rm hor}$, the proper length from the inner horizon to the 
singularity.

\subsection{One point function   $\langle O^2 \rangle $ from CFT}

From the analysis of the rotating BTZ background we saw that in the large horizon radius limit, or the large spatial 
length limit, the leading contribution to the thermal one point function does not see $\ell_{\rm sing}$, 
the geometric length
from the inner horizon to the singularity.
We know that composite operators, or operators which do not transform as conformal primaries gain thermal expectation 
value in 2d CFT on a plane.
In this section we evaluate this one point function and observe that when their conformal dimensions are large, 
their expectation value can be cast
into the geometric form, but again there is no information about $\ell_{\rm sing}$. 

\begin{figure}[t]
	\centering
	\includegraphics[scale=.5]{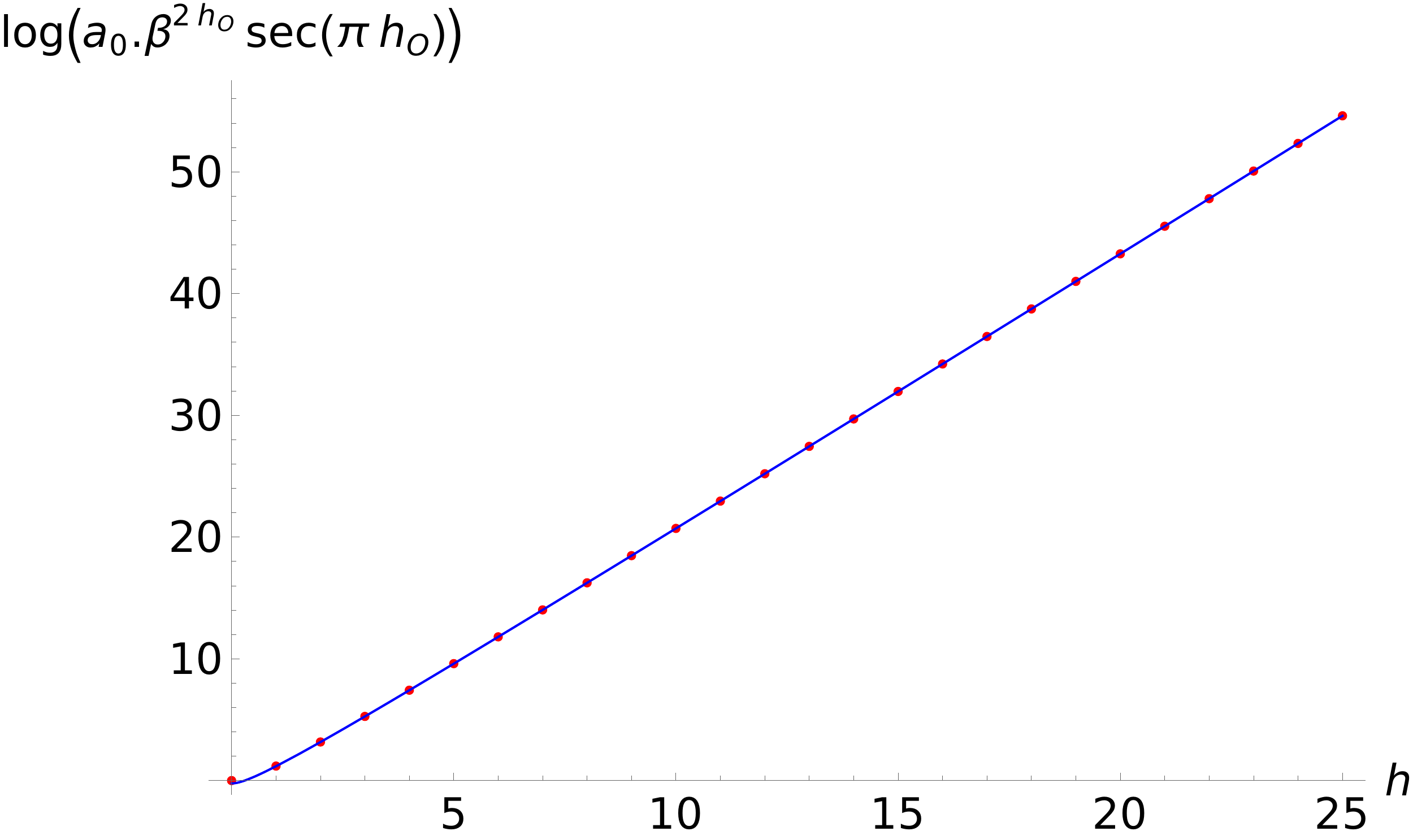}
	\caption{The blue curve represents the plot of the function $ \log[a_0 \beta^{2 h_\mathcal{O}}\sec(\pi h_\mathcal{O})] $, where $ a_0 $ is obtained from the result of our saddle point analysis given in the equation \eqref{a_0}.  The red points are the exact values of the $ \log[a_0 \beta^{2 h_\mathcal{O}} (-1)^{h_{\cal O}}] $ at the integer values of $ h_\mathcal{O} $ obtained by explicitly 
evaluating the residue
	\eqref{residue}. 
	 }
	\label{fig:1}
\end{figure}

We consider the bilinear compose ${\cal O}^2 $ of the primary ${\cal O}$ of dimension  $2h_{\cal O}$ and let $2h_{\cal O}$ 
be an integer.  Here the holomorphic and the anti-holomorphic dimension of the operator is $(h_{\cal O}, h_{\cal O})$. 
The  one point function of the operator of the composite can be obtained by  the following approach. 
We first look at the 2 point function  at finite temperature
\begin{align} \label{cftcor}
	\langle \mathcal{O}(w,\bar{w})\mathcal{O}(0, 0)\rangle&=\bigg[\frac{\beta_L}{\pi}\sinh\bigg(\frac{\pi w}{\beta_L}\bigg)\bigg]^{-2h_{\mathcal{O}}}\bigg[\frac{\beta_R}{\pi}\sinh\bigg(\frac{\pi \bar{w}}{\beta_R}\bigg)\bigg]^{-2h_{\mathcal{O}}} .
\end{align}
Here the left and right inverse temperatures of the CFT is given by $(\beta_L, \beta_R)$ and since we are 
interested in the BTZ black hole with angular momentum we should have $\beta_L \neq \beta_R$. 
The  expectation value $\langle {\cal O}^2 \rangle$  can be obtained by examining  the  expansion in $w, \bar w$ of 
(\ref{cftcor}) and extracting out the constant term. 
To see that this indeed captures thermal expectation value, 
we know that the OPE of ${\cal O}$ with itself contains the composite ${\cal O}^2$. 
\begin{equation}
{\cal O}( z, \bar z) {\cal O}(0, 0)  \sim \frac{1}{|z|^{4h_{\cal O}}} + \cdots  +  C  {\cal O}^2(0, 0) + 
\cdots \ .
\end{equation}
Here the $\cdots$ refer to 
other  singular terms or terms which depend on powers of $z, \bar z$  in the OPE which we are not interested in. 
The ${\cal O}^2$  operator occurs in the OPE with no 
dependence of $z$ and  $C$ is the OPE coefficient. 
Now taking thermal expectation value on both sides of this equation, we see that the constant term, independent 
of $z, \bar z$ is proportional to the expectation value  $\langle {\cal O}^2 \rangle$. 
To extract the constant term, it is sufficient to look at the holomorphic two point function and obtain the residue in the Laurent expansion in 
$z$. 
The residue when  $2h_{\cal O}$ is an integer is given by 
\begin{equation} \label{residue}
a_0 (\beta) = 
\frac{1}{2\pi i}\oint \frac{dw}{w}\bigg[\frac{\beta}{\pi}\sinh\bigg(\frac{\pi w}{\beta}\bigg)\bigg]^{-2 h_{\mathcal{O}}} ,
\end{equation}
where the integral is around the origin. 
Then from the OPE expansion of (\ref{cftcor}), the thermal expectation value is given by 
\begin{equation}
C\langle {\cal O } ^2\rangle = a_0( \beta_L) a_0(\beta_R) \,.
\end{equation}
Examining the residue (\ref{residue}), we see that it is non-zero only when $2h_0$ is an even integer or $h_{\cal O}$ is an integer.
We re-write the residue as 
\begin{equation} \label{saddle}
a_0( \beta) = \frac{1}{2\pi i } \Big( \frac{\pi}{\beta} \Big)^{2h_{\cal O} } \oint dy e^{-\left[  \log y + 2h_{\cal O} \log( \sinh y) \right]} \,.
\end{equation}
For large $2h_{\cal O}$ we can perform this integral by the saddle point method. 
The saddle points  at the leading order are at 
\begin{equation}
y_{\rm saddle}  = \pm \frac{ i \pi }{2}  + O(\frac{1}{h_{\cal O}} ) \,.
\end{equation}
Adding the contributions at the 2 saddles along with the one loop term at each of the saddles we obtain 
\begin{align}\label{a_0}
		a_0 (\beta)=\frac{1}{2\pi i}\oint \frac{dw}{w}\bigg(\frac{\beta}{\pi}\sinh\big[\frac{\pi w}{\beta}\big]\bigg)^{-2 h_{\mathcal{O}}}\approx\frac{2\pi^{2h_\mathcal{O}-\frac{1}{2}}\cos(\pi h_{\mathcal{O}})}{\beta^{2h_\mathcal{O}}\ \sqrt{2+\pi^2h_\mathcal{O}}} \,.
\end{align}
We can compare the value of $a_0$ from the saddle point against the numerical value, which can be obtained by 
explicitly evaluating the residue. This comparison is shown in the table \ref{table1}  as well as in the graph in figure 
\ref{fig:1}, 
we see that indeed, that the saddle point  approximation is a good approximation at large values of integer $2h_{\cal O}$.
\begin{table}[ht]
\centering{ \footnotesize{
	\begin{tabular}{|c | c | c |c| }
			\hline
			$  $& $   $&$  $& $  $ \\
			\parbox{3cm}{\centering  {{$ h_\mathcal{O} $} }} &{$ (a_0\beta^{2h_\mathcal{O}}\pi^{\frac{1}{2}-2h_\mathcal{O}}) $}  Exact & \parbox{5cm}{\centering {$ (a_0\beta^{2h_\mathcal{O}}\pi^{\frac{1}{2}-2h_\mathcal{O}} )$} Saddle point} & Error in $ \% $  \\ 
			$  $& $   $&$  $& $  $ \\
			\hline
			$  $& $   $&$  $& $  $ \\
				$ 5 $ & $ -0.280338 $ & -0.279105 & 0.439624 \\
				$ 10 $ & $ 0.199763 $ & $ 0.199308 $ & $ 0.227946 $ \\
				$ 15 $ & $ -0.163527 $ & $ -0.163275 $ & $ 0.153875 $ \\
				$ 20 $ & 0.141801 & $ 0.141637 $ & $ 0.116138 $ \\
				$ 25 $ & $ -0.126929 $ & $ -0.126811 $ & $ 0.0932657 $ \\
				$ 30$ &$ 0.11593 $& $0.11584 $& $0.07792 $ \\
				$ 35 $ & $ -0.10737 $ & $ -0.107298 $ & $ 0.0669107 $ \\
				$ 40 $ & $ 0.100463 $ & $ 0.100404 $ & $ 0.0586273 $ \\
				$ 45 $ & $ -0.0947381 $ & $ -0.0946887 $ & $ 0.052169 $ \\
				$ 50 $ & $ 0.089892 $ & $ 0.0898497 $ & $ 0.0469923 $ \\
				$ 55 $ & $ -0.0857207 $ & $ -0.0856841 $ & $ 0.0427502 $ \\
				$ 60 $ & $ 0.082081 $ & $ 0.0820488 $ & $ 0.0392106 $ \\
				$ 65 $ & $ -0.0788687 $ & $ -0.0788402 $ & $ 0.0362124 $ \\ 
				$  $& $   $&$  $& $  $ \\ \hline
	\end{tabular}  
	\caption{Numerical comparison of the exact value of the residue at integer $h_{\cal O}$ against the saddle 
	point given in (\ref{a_0}).} \label{table1}
}}
\end{table}

From the approximation of $a_0(\beta)$ in (\ref{a_0}), we see  that the thermal one point function is given by 
in the large $2h_{\cal O}$ limit is given by
\begin{eqnarray} \label{2oexp}
 \langle {\cal O }^2 \rangle &=&  \frac{1}{C } a_0(\beta_L) a_0(\beta_R) \,, \\ \nonumber
 &=&   \frac{1}{C  } 
  \bigg(\frac{1}{\beta_L\beta_R}\bigg)^{2 h_\mathcal{O}} \ \frac{4\pi^{4h_\mathcal{O}-1}\cos^2(\pi h_{\mathcal{O}})}{\ {2+\pi^2h_\mathcal{O}}}\,.
\end{eqnarray}
The $\cos^2(\pi h_{\mathcal{O}})$ term is just unity when  $h_{{\cal O}}$ is an integer, but we keep this
factor so that it enables us to obtain the general form of the result. 
Let us now recast this expression in terms of geometric quantities in the bulk. 
The left and right inverse temperatures are related to the horizon radii by  \cite{Maldacena:1998bw}
\begin{equation}
\frac{1}{\beta_L} = \frac{r_+ - r_-}{2\pi R^2} \,, \qquad  \frac{1}{\beta_L} = \frac{r_+ +  r_-}{2\pi R^2} \,.
\end{equation}
Using this relation in (\ref{2oexp}), we can write 
\begin{equation} \label{btzholimit}
\lim_{h_{\rm O} \rightarrow \infty + i \epsilon} 
 \langle {\cal O }^2 \rangle \sim  \frac{1}{C z_0^{4 h_{\cal O} } } 
 e^{ -2  h_{\cal O} \log  4 +2 h_{\cal O}  \log( 1- \delta^2) }
 \exp ( - i 2 \pi  h_{\cal O}   )  \,.
 \end{equation}
The small  imaginary part selects the leading phase  to be $ \exp ( -2 \pi i  h_{\cal O}  ) $. 
In gravity the $3$ OPE coefficient is given by $C=  \sqrt{2} $  \cite{Belin:2017nze}, 
 therefore they  does not grow exponentially 
in $h_{\cal O}$. 
We can also identify $2h_{\cal O} = \frac{\Delta_{{\cal O}^2}}{2}$. 
With these inputs, the thermal one point function can be written as  
\begin{eqnarray}
 \langle {\cal O }^2 \rangle &\sim &
 \exp\Big[  -i \pi  \frac{ \Delta_{{\cal O}^2} }{2 }  -  \frac{ \Delta_{{\cal O}^2} }{2 }  \big( \log4 -   \log( 1- \delta^2) \big )  \Big] ,
 \\ \nonumber
 &\sim& \exp\big[ -i \frac{\pi 2 m R}{2}  - \frac{2m R}{2}\big(  \log 4  -    \log( 1- \delta^2) \big)  \Big].
\end{eqnarray}
Using (\ref{btztaus}), (\ref{lhorbtz}), we   can write the expectation value of the composite as
\begin{equation}
\langle {\cal O }^2 \rangle   \sim e^{ -i 2m \tau_s  - 2m \ell_{\rm hor} } .
\end{equation}
Note that $m\rightarrow 2m$ since we are evaluating the expectation value of the composite ${\cal O}^2$, 
and we do not see the appearance of $\ell_{\rm sing}$.

Though our calculation in the CFT has been done only for integer conformal dimensions, we have written the result
so that it is natural to read out the general form of the result.  But, it will be interesting to see if the methods of \cite{Iliesiu:2018fao}
can be used to verify if indeed the expectation value of the composite can be written in the form 
for arbitrary conformal dimensions \footnote{ In \cite{Iliesiu:2018fao} 
the  expectation value of the composite in which the operator ${\cal O}$ obeys mean field theory 
correlators have been written for 2d CFT, but  not for the correlator given in (\ref{cftcor}).  We have verified that
the expectation value of the composite ${\cal O}^2$ using the mean field theory correlator does not 
behave as in (\ref{2oexp}) in the large $h_{\cal O}$ limit.   }. 

\section{Conclusions}
\label{section5}

We have discussed two examples in which the thermal one point function of massive scalars can be evaluated exactly. 
For the charged planar black hole we introduced a large $d$ limit to achieve this. 
In both these examples the result for the one point function agreed with that 
anticipated in \cite{Grinberg:2020fdj} using WKB methods. 
In particular for the charged black hole which has an inner horizon we observed the presence 
of $\ell_{hor}$, the proper distance from the inner horizon to the singularity. 
The second example involved hyperbolic black holes for which $\tau_s$, the time to the singularity has a different 
dependence on the radius of $AdS_{d+1}$. 
We also saw that these results remain the same for both the Gauss-Bonnet or the Weyl tensor squared coupling. 

We obtained the one point function of massive scalars in BTZ due to a cubic coupling of scalars. 
We could interpret this result geometrically in a suitable limit of large horizon radius, but 
could not find the dependence of $\ell_{\rm sing}$ in this limit. 
It would be interesting to investigate the case of rotating black hole in higher dimensions. 
The WKB methods of \cite{Grinberg:2020fdj}  assumed radial symmetry and it would be important to generalize 
those arguments to backgrounds with axial symmetry.  The recent developments
\cite{Dodelson:2022yvn,Bhatta:2022wga} regarding the applications of 
 the solutions to the Heun equation to evaluate thermal 2 point functions  will be helpful in this regard. 

We saw that  in $2d$ we could gain some insight on thermal one point functions by evaluating the 
expectation value of composite scalars at finite temperature. 
For this we used integer conformal dimensions. 
The methods of \cite{Iliesiu:2018fao} can be used to study this question for arbitrary conformal dimensions.
It will be important to develop such methods in field theory to understand the features of thermal 
one point function. 
As we have seen, these contain some information of the geometry behind the horizon and evaluating one
thermal one point functions directly in the dual field theory can give us new insight to black hole geometry.

 \acknowledgments
We thank Chethan Krishnan, Suvrat Raju and Aninda Sinha for discussions during the presentation of a preliminary version of this work.   J.R.D thanks the group at the Asia Pacific Center for Theoretical Physics, Pohang for the 
warm hospitality during the completion of this work. 

\appendix
\section{Details for the Green's function}

\label{appendixa}

The bulk-bulk Green's function $G(z, z')$ satisfies the equation,
\begin{equation}
\partial_z\left( \frac{R^{d-1}}{z^{d-1} } f(z)  \partial_z G(z, z') \right)  - \frac{R^{d-1}}{z^{d-1}} m^2 G(z, z')  = 
\delta( z-z') \,, \label{A.1}
\end{equation}
 and $ G(z,z') $ has to be well-behaved at the boundaries  $ z=0 $ and $ z=z_0 $; $ f(z) $ was defined in the equation \eqref{adsbh}. \\

Now, the two linearly independent solutions of the homogeneous part of the differential equation \eqref{A.1} are given by,
\begin{align}
	g_{{\rm inf} }(w)=w^h\ {}_2F_1(h,h,2h,w) \,,\qquad\text{and,} \qquad  g_{{\rm hor} }(w)= w^h {}_2F_1(h,h,1,1-w) \,.
\end{align}
\begin{align}
\text{Where, }\qquad w = \left( \frac{z}{z_0}\right)^d,  \qquad\qquad\qquad h = \frac{\Delta}{d} \,, 
\end{align}
The solutions $ g_{\rm inf}(w) $ and $ g_{\rm hor}(w) $ are regular at $ w=0 $ and at $ w=1 $ respectively, which indicates,
	\begin{align}\begin{split}
G(w,w')=\begin{cases}
A(w') g_{\rm inf}(w)\ \ \text{for}\  w<w' ,\\
B(w') g_{\rm hor}(w)\ \ \text{for}\ w>w' .\\
\end{cases}
\end{split}\end{align}
Where $ A(w') $ and $ B(w') $ are functions of $ w' $, to be determined.
\\

Now, for the Green's function to be continuous at $ w=w' $, we should have,
\begin{align}\label{A.5}
A(w') g_{\rm inf}(w')=B(w') g_{\rm hor}(w') \,. \end{align}

	The jump of $ \partial_wG(w,w') $ around $ w=w' $ is obtained by integrating the both side of the equation \eqref{A.1} with respect to $ z $ from $ z=z'-\epsilon $ to $ z=z'+\epsilon $ with $ \epsilon\rightarrow 0 $ and then we used the equation \eqref{A.5} to get,

	\begin{align}\label{1.0.25}
	\frac{B(w')}{g_{\rm inf}(w')}\bigg(g_{\rm inf}(w')g_{hor}'(w')-g_{hor}(w')g_{\rm inf}'(w')\bigg)=\frac{z_0^d}{R^{d-1}d(1-w)} \,.
	\end{align}
	Note that the terms inside the parenthesis in the l.h.s is equal to the Wronskian of the  homogeneous part of the differential equation \eqref{A.1} written in $ w' $ variable. And this Wronskian is calculated up to an overall constant factor $ \tilde{C} $ directly from the differential equation, thus we have,
	\begin{align}
	\bigg(g_{\rm inf}(w')g_{hor}'(w')-g_{\rm hor}(w')g_{\rm inf}'(w')\bigg)=\frac{\tilde{C}}{1-w'} \,.
	\end{align}
	Now, the constant $ \tilde{C} $ is determined by expanding the both side of the above equation in the taylor series about $ w=0 $. The first term of the series expansion of the l.h.s is equal to $ \tilde{C} $. Thus one obtains,
	\begin{align}
	\tilde{C}=-\frac{\Gamma(2h)}{\Gamma(h)^2} \,.
	\end{align}
	Hence,
	\begin{align}
	B(w')=-\frac{\Gamma(h)^2}{\Gamma(2h)}\ \frac{z_0^d}{R^{d-1}d}\ g_{\rm inf}(w')\ \ , \qquad\qquad A(w')=-\frac{\Gamma(h)^2}{\Gamma(2h)}\ \frac{z_0^d}{R^{d-1}d}\ g_{\rm hor}(w') \,.
	\end{align}
	And finally, the bulk-bulk green's function,
	\begin{align}
	G(w,w')=-\frac{\Gamma(h)^2}{\Gamma(2h)}\ \frac{z_0^d}{R^{d-1}d}\bigg(g_{{\rm inf} }(w)g_{{\rm hor} }(w')\theta(w'-w)+g_{{\rm hor} }(w)g_{{\rm inf} }(w')\theta(w-w')\bigg) \,. \nonumber
	\end{align}
	
	\section{The charged planar black hole solution  at large $ d $}	\label{appendixb}

	The consistency of the treatment of large $ d $ limit used to evaluate the one point function for the case of the charged planar black hole requires the metric \eqref{chargmet} in this limit to satisfy the Einstien-Maxwell field equation at the leading order in $ d $. In this appendix, we will verify it by straightforward calculation.
	
	At the limit described in \eqref{dinfty}, the metric \eqref{chargmet} takes the form,
	\begin{align}\label{metdinfty}
		ds^2\approx\frac{R_{AdS}^2}{z^2}\bigg(-\hat{f}(z)dt^2+\frac{dz^2}{\hat{f}(z)}+d\vec{x}^2\bigg),\ \ \text{where}\ \ \ \ \ \hat{f}(z)=1-\frac{z^d}{z_0^d}+Q\frac{z^{2d}}{z_0^{2d}} \,.
	\end{align}
	Note that the asymptotic behaviour of the metric at the boundary is kept intact in this limit, thus the rules of AdS/CFT were implemented to obtain the one-point function.\\\\
	The above metric should satisfy the following Einstien-Maxwell field equation at large $ d $,
	\begin{align}\label{Ein-Max}
	\lim_{d\rightarrow\infty}\bigg(R_{\mu\nu}-\frac{1}{2}g_{\mu\nu}R-\Lambda g_{\mu\nu}\bigg)=-\lim_{d\rightarrow\infty}8\pi T_{\mu\nu} \,,
	\end{align}
	where the cosmological constant $ \Lambda=-\frac{d(d-1)}{2R_{AdS}^2} $ for the $ AdS_{d+1} $ and $ T_{\mu\nu} $ is the energy-momentum stress tensor evaluated from the gauge potential \cite{Chamblin:1999tk},
	\begin{align}
		A=\bigg(-\frac{q z^{d-2}}{c}+\Phi\bigg)dt,  \qquad	\Phi\ \text{is constant, and }\quad  c=\sqrt{\frac{2(d-2)}{d-1}} \,.
	\end{align}
	By calculating the l.h.s and r.h.s separately for each tensorial component of the equation \eqref{Ein-Max}, after incorporating the metric \eqref{metdinfty} in it, we will show that l.h.s$ = $r.h.s for each component. For this calculation we need the following results,

	\begin{align}
		&R_{00}=\frac{f(z) (-2f(z)d+z[(1+d)f'(z)-zf''(z)])}{2 z^2} \,,\\
		&R_{11}=\frac{2f(z)d+z[-(d+1)f'(z)+zf''(z)]}{2z^2f(z)}\,,\\
		&R_{ii}=\frac{df(z)-zf'(z)}{z^2}\ \ \text{for}\ \ i\in[2,d] \,.
	\end{align}
	All other components of Ricci-tensor are zero. And the Ricci-scalar is given by,
	\begin{align}
		R=d(d+1)f(z)+z(-2f'(z)d+zf''(z)) \,.
	\end{align}
	For the 00-component,
	\begin{align}
		\lim_{d\rightarrow\infty}\bigg(R_{00}-\frac{1}{2}g_{00}R-\Lambda g_{00}\bigg)&=\lim_{d\rightarrow\infty}\bigg(-\frac{d(d-1)}{2}Qz^{2d-2}z_0^{-4d}\big[Qz^{2d}+z_0^d(-z^d+z_0^d)\big]\bigg) , \nonumber\\
		&=\frac{1}{z^2}\bigg[-e^{2d\log(\frac{z}{z_0})}+e^{3d\log(\frac{z}{z_0})}-e^{4d\log(\frac{z}{z_0})}\bigg] .
	\end{align}
	\begin{align}
		-\lim_{d\rightarrow\infty}8\pi T_{00}&=-\lim_{d\rightarrow\infty}\bigg(\frac{(d-2)(d-1)}{2}Qz^{2d-4}z_0^{2-4d}[Qz^{2d}+z_0^d(-z^d+z_0^d)]\bigg) ,\\
		&=\frac{1}{z^2}\bigg[-e^{2d\log(\frac{z}{z_0})}+e^{3d\log(\frac{z}{z_0})}-e^{4d\log(\frac{z}{z_0})}\bigg] .
	\end{align}
	For the 11-component,
	\begin{align}
		\lim_{d\rightarrow\infty}\bigg(R_{11}-\frac{1}{2}g_{11}R-\Lambda g_{11}\bigg)&=\lim_{d\rightarrow\infty}\frac{d(d-1)Qz^{2d-2}}{2(Qz^{2d}+z_0^2d-z^dz_0^d)} \,,\\
		&=\frac{1}{z^2}\bigg(\frac{1}{e^{2d\log(\frac{z_0}{z})}-e^{d\log(\frac{z_0}{z})}+Q}\bigg) .
	\end{align}
	\begin{align}
		-\lim_{d\rightarrow\infty}8\pi T_{11}&=\lim_{d\rightarrow\infty}\bigg(\frac{(d-2)(d-1)Qz_0^2}{2z^4[Q+z^{-2d}z_0^d(-z^d+z_0^d)]}\bigg) \,,\\
		&=\frac{1}{z^2}\bigg(\frac{1}{e^{2d\log(\frac{z_0}{z})}-e^{d\log(\frac{z_0}{z})}+Q}\bigg) .
	\end{align}
	and, for $ ii $-component,
	\begin{align}
		\lim_{d\rightarrow\infty}\bigg(R_{ii}-\frac{1}{2}g_{ii}R-\Lambda g_{ii}\bigg)&=-\lim_{d\rightarrow\infty}\bigg(\frac{d(d+1)Qz^{2d-2}z_0^{-2d}}{2}\bigg) ,\\
		&=-\frac{1}{z^2}e^{2d\log(\frac{z}{z_0})} .
	\end{align}
	\begin{align}
		-\lim_{d\rightarrow\infty}8\pi T_{ii}&=\lim_{d\rightarrow\infty}\bigg(-\frac{(d-2)(d-1)Qz^{2d-4}z_0^{2-2d}}{2}\bigg) ,\\
		&=-\frac{1}{z^2}e^{2d\log(\frac{z}{z_0})}.
	\end{align}
	Thus, for each tensorial component we have shown that l.h.s$ = $r.h.s, i.e., the  Einstien-Maxwell field equation is satisfied by the metric given in \eqref{metdinfty} at the leading order in $ d $.

\section{Large $d$ with  Weyl tensor squared coupling }
 \label{appendixc}
 
	We repeat the calculation for the one point function in the charged planar black hole background at large $ d $, described in section \ref{section3}, but using the Weyl tensor squared coupling in place of the GB coupling.\\
	
	The rules of AdS/CFT prescribe the one point function to be,
	\begin{equation} \label{1ptgenW}
		\langle O(t, \vec x ) \rangle  = \alpha \int dz dt' d\vec x'
		\sqrt{g} \tilde K( t, \vec x; \,  z, t', \vec x') {W^2}_{} (z, t', \vec x' ) ,
	\end{equation}
	where the $ W^2=W_{\mu\nu\rho\sigma}W^{\mu\nu\rho\sigma} $, and it would be sufficient to do only the $ z $ part of the above integral as was argued before.\\
	
	For the metric given in \eqref{chargmet},
	\begin{align}
		W^2=\frac{(d-2)(d-1)^2w^{2-\frac{4}{d}}[(6-4d)Qw+dw^{2/d}]^2}{R^4d} \, .
	\end{align}
	And in the limit $ d\rightarrow\infty $, the leading contribution to the Weyl tensor's square is given by,
	\begin{align}
		W^2=\frac{w^2 (1-4Qw)^2d^4}{R^4} \,.
	\end{align}
	Now plugging the expression for the bulk-boundary green's function $ K(w) $, given in \eqref{bbpropc}, in the equation \eqref{1ptgenW}, and using the change of variable defined in \eqref{changeofvari}, we get,
	\begin{align}\begin{split}
	\langle O\rangle=&-\frac{4d^2(1-4Q)^{h/2}}{R^2}\frac{(\Gamma(h))^2}{\Gamma(2h)} \bigg(\frac{2\nu }{z_0^{\Delta_+}}\bigg)\int_{0}^{1}dy(1-y)^h{}_2F_1(h,h,1,y)\times\\&
	\bigg[(2Qw_+)^{-2}(1-\chi y)^{-2}-16Q(2Qw_+)^{-3}(1-y)(1-\chi y)^{-3}+64Q^2(2Qw_+)^{-4}(1-y)^2(1-\chi y)^{-4}\bigg] .
	\end{split}\end{align}
	Applying the integration formulae used in the section \ref{section3}, the integrals in the above equation are performed to get the analytic expression for the thermal one point function, 
	\begin{align}\begin{split}
	\langle O\rangle=&-\frac{4d^2(1-4Q)^{\frac{h}{2}-1}}{R^2}\frac{(\Gamma(h))^2}{\Gamma(2h)}
	\frac{h(h-1)\pi\csc(\pi h)}{12(1-\sqrt{1-4Q}) }\bigg(\frac{2\nu }{z_0^{\Delta_+}}\bigg)\times\\
	&\bigg(\big[3(-1+\sqrt{1-4Q})+8Q(h-2)\{h(-1+\sqrt{1-4Q})+\sqrt{1-4Q}\}\big]\\&\hspace{4cm}\times{}_2F_1(2-h,1+h,2,\frac{1}{2}-\frac{1}{2\sqrt{1-4Q}})\\
	&\hspace{3cm}-8Q(h-2)\ {}_2F_1(3-h,1+h,2,\frac{1}{2}-\frac{1}{2\sqrt{1-4Q}})\bigg) \,.
	\end{split}\end{align}
	An important point to note that this Weyl tensor square induced one point function differs from the one point function enabled by the GB coupling given in \eqref{finachargr}, by the polynomial factors in front of the hypergeometric functions which are not so crucial in the large $ h $ bahaviour of the one point function as the more dominant contributions come from the exponential growth of the hypergeometric functions at large $ h $.\\
	
	Thus, the one point function undergoes the exact similar treatment of taking the large $ h $ limit as was illustrated in section \ref{section3} to land up into the form,
	\begin{align}
	\langle O\rangle\sim \exp\bigg[m\bigg({{-\frac{R}{d}\log\bigg[\frac{4}{\sqrt{1-4Q}}\bigg]}+{\frac{R}{d}\arcsech(\sqrt{1-4Q})}}-i{\pi \frac{R}{d}}\bigg)\bigg] .
	\end{align}
	So, we observe that the thermal one point function due to both the Weyl tensor squared and GB coupling behaves identically at large $ h $.

\bibliographystyle{JHEP}
\bibliography{references} 

\providecommand{\href}[2]{#2}\begingroup\raggedright\begin{thebibliography}{10}

\bibitem{Maldacena:1997re}
J.~M. Maldacena, {\it {The Large N limit of superconformal field theories and
  supergravity}},  {\em Adv. Theor. Math. Phys.} {\bf 2} (1998) 231--252,
  [\href{http://arxiv.org/abs/hep-th/9711200}{{\tt hep-th/9711200}}].

\bibitem{Fidkowski:2003nf}
L.~Fidkowski, V.~Hubeny, M.~Kleban, and S.~Shenker, {\it {The Black hole
  singularity in AdS / CFT}},  {\em JHEP} {\bf 02} (2004) 014,
  [\href{http://arxiv.org/abs/hep-th/0306170}{{\tt hep-th/0306170}}].

\bibitem{Grinberg:2020fdj}
M.~Grinberg and J.~Maldacena, {\it {Proper time to the black hole singularity
  from thermal one-point functions}},  {\em JHEP} {\bf 03} (2021) 131,
  [\href{http://arxiv.org/abs/2011.01004}{{\tt arXiv:2011.01004}}].

\bibitem{Rodriguez-Gomez:2021pfh}
D.~Rodriguez-Gomez and J.~G. Russo, {\it {Correlation functions in finite
  temperature CFT and black hole singularities}},  {\em JHEP} {\bf 06} (2021)
  048, [\href{http://arxiv.org/abs/2102.11891}{{\tt arXiv:2102.11891}}].

\bibitem{McInnes:2022pig}
B.~McInnes, {\it {Inside Flat Event Horizons}},
  \href{http://arxiv.org/abs/2206.00198}{{\tt arXiv:2206.00198}}.

\bibitem{Georgiou:2022ekc}
G.~Georgiou and D.~Zoakos, {\it {Holographic correlation functions at finite
  density and/or finite temperature}},  {\em JHEP} {\bf 11} (2022) 087,
  [\href{http://arxiv.org/abs/2209.14661}{{\tt arXiv:2209.14661}}].

\bibitem{Berenstein:2022nlj}
D.~Berenstein and R.~Mancilla, {\it {Aspects of thermal one-point functions and
  response functions in AdS Black holes}},
  \href{http://arxiv.org/abs/2211.05144}{{\tt arXiv:2211.05144}}.

\bibitem{David:2011hy}
J.~R. David, S.~Jain, and S.~Thakur, {\it {Shear sum rules at finite chemical
  potential}},  {\em JHEP} {\bf 03} (2012) 074,
  [\href{http://arxiv.org/abs/1109.4072}{{\tt arXiv:1109.4072}}].

\bibitem{David:2012cd}
J.~R. David and S.~Thakur, {\it {Sum rules and three point functions}},  {\em
  JHEP} {\bf 11} (2012) 038, [\href{http://arxiv.org/abs/1207.3912}{{\tt
  arXiv:1207.3912}}].

\bibitem{Myers:2016wsu}
R.~C. Myers, T.~Sierens, and W.~Witczak-Krempa, {\it {A Holographic Model for
  Quantum Critical Responses}},  {\em JHEP} {\bf 05} (2016) 073,
  [\href{http://arxiv.org/abs/1602.05599}{{\tt arXiv:1602.05599}}]. [Addendum:
  JHEP 09, 066 (2016)].

\bibitem{Emparan:2020inr}
R.~Emparan and C.~P. Herzog, {\it {Large D limit of Einstein\textquoteright{}s
  equations}},  {\em Rev. Mod. Phys.} {\bf 92} (2020), no.~4 045005,
  [\href{http://arxiv.org/abs/2003.11394}{{\tt arXiv:2003.11394}}].

\bibitem{Giataganas:2021jbj}
D.~Giataganas, N.~Pappas, and N.~Toumbas, {\it {Holographic observables at
  large d}},  {\em Phys. Rev. D} {\bf 105} (2022), no.~2 026016,
  [\href{http://arxiv.org/abs/2110.14606}{{\tt arXiv:2110.14606}}].

\bibitem{Kraus:2016nwo}
P.~Kraus and A.~Maloney, {\it {A cardy formula for three-point coefficients or
  how the black hole got its spots}},  {\em JHEP} {\bf 05} (2017) 160,
  [\href{http://arxiv.org/abs/1608.03284}{{\tt arXiv:1608.03284}}].

\bibitem{Witten:1998qj}
E.~Witten, {\it {Anti-de Sitter space and holography}},  {\em Adv. Theor. Math.
  Phys.} {\bf 2} (1998) 253--291,
  [\href{http://arxiv.org/abs/hep-th/9802150}{{\tt hep-th/9802150}}].

\bibitem{Sen:2005iz}
A.~Sen, {\it {Entropy function for heterotic black holes}},  {\em JHEP} {\bf
  03} (2006) 008, [\href{http://arxiv.org/abs/hep-th/0508042}{{\tt
  hep-th/0508042}}].

\bibitem{Banks:1998dd}
T.~Banks, M.~R. Douglas, G.~T. Horowitz, and E.~J. Martinec, {\it {AdS dynamics
  from conformal field theory}},
  \href{http://arxiv.org/abs/hep-th/9808016}{{\tt hep-th/9808016}}.

\bibitem{Erbin}
H.~Erbin, {\it {Scalar propagators on adS space}}, 
  \url{https://www.lpthe.jussieu.fr/~erbin/files/ads_propagators.pdf}.

\bibitem{Chamblin:1999tk}
A.~Chamblin, R.~Emparan, C.~V. Johnson, and R.~C. Myers, {\it {Charged AdS
  black holes and catastrophic holography}},  {\em Phys. Rev. D} {\bf 60}
  (1999) 064018, [\href{http://arxiv.org/abs/hep-th/9902170}{{\tt
  hep-th/9902170}}].

\bibitem{gradshteyn2007}
I.~S. Gradshteyn and I.~M. Ryzhik, {\em Table of integrals, series, and
  products}.
\newblock Elsevier/Academic Press, Amsterdam, seventh~ed., 2007.
\newblock Translated from the Russian, Translation edited and with a preface by
  Alan Jeffrey and Daniel Zwillinger, With one CD-ROM (Windows, Macintosh and
  UNIX).

\bibitem{Watson}
G.~N. Watson, {\it {Asymptotic expansion of hypergeometric functions}},  {\em
  Trans. Cambridge. Philos. Soc.} {\bf 22} (1918) 277--308.

\bibitem{Emparan:1998he}
R.~Emparan, {\it {AdS membranes wrapped on surfaces of arbitrary genus}},  {\em
  Phys. Lett. B} {\bf 432} (1998) 74--82,
  [\href{http://arxiv.org/abs/hep-th/9804031}{{\tt hep-th/9804031}}].

\bibitem{Birmingham:1998nr}
D.~Birmingham, {\it {Topological black holes in Anti-de Sitter space}},  {\em
  Class. Quant. Grav.} {\bf 16} (1999) 1197--1205,
  [\href{http://arxiv.org/abs/hep-th/9808032}{{\tt hep-th/9808032}}].

\bibitem{Emparan:1999gf}
R.~Emparan, {\it {AdS / CFT duals of topological black holes and the entropy of
  zero energy states}},  {\em JHEP} {\bf 06} (1999) 036,
  [\href{http://arxiv.org/abs/hep-th/9906040}{{\tt hep-th/9906040}}].

\bibitem{Casini:2011kv}
H.~Casini, M.~Huerta, and R.~C. Myers, {\it {Towards a derivation of
  holographic entanglement entropy}},  {\em JHEP} {\bf 05} (2011) 036,
  [\href{http://arxiv.org/abs/1102.0440}{{\tt arXiv:1102.0440}}].

\bibitem{Camporesi:1994ga}
R.~Camporesi and A.~Higuchi, {\it {Spectral functions and zeta functions in
  hyperbolic spaces}},  {\em J. Math. Phys.} {\bf 35} (1994) 4217--4246.

\bibitem{Banados:1992gq}
M.~Banados, M.~Henneaux, C.~Teitelboim, and J.~Zanelli, {\it {Geometry of the
  (2+1) black hole}},  {\em Phys. Rev. D} {\bf 48} (1993) 1506--1525,
  [\href{http://arxiv.org/abs/gr-qc/9302012}{{\tt gr-qc/9302012}}]. [Erratum:
  Phys.Rev.D 88, 069902 (2013)].

\bibitem{DHoker:1999mqo}
E.~D'Hoker, D.~Z. Freedman, and L.~Rastelli, {\it {AdS / CFT four point
  functions: How to succeed at z integrals without really trying}},  {\em Nucl.
  Phys. B} {\bf 562} (1999) 395--411,
  [\href{http://arxiv.org/abs/hep-th/9905049}{{\tt hep-th/9905049}}].

\bibitem{Maldacena:1998bw}
J.~M. Maldacena and A.~Strominger, {\it {AdS(3) black holes and a stringy
  exclusion principle}},  {\em JHEP} {\bf 12} (1998) 005,
  [\href{http://arxiv.org/abs/hep-th/9804085}{{\tt hep-th/9804085}}].

\bibitem{Belin:2017nze}
A.~Belin, C.~A. Keller, and I.~G. Zadeh, {\it {Genus two partition functions
  and R\'enyi entropies of large c conformal field theories}},  {\em J. Phys.
  A} {\bf 50} (2017), no.~43 435401,
  [\href{http://arxiv.org/abs/1704.08250}{{\tt arXiv:1704.08250}}].

\bibitem{Iliesiu:2018fao}
L.~Iliesiu, M.~Kolo\u{g}lu, R.~Mahajan, E.~Perlmutter, and D.~Simmons-Duffin,
  {\it {The Conformal Bootstrap at Finite Temperature}},  {\em JHEP} {\bf 10}
  (2018) 070, [\href{http://arxiv.org/abs/1802.10266}{{\tt arXiv:1802.10266}}].

\bibitem{Dodelson:2022yvn}
M.~Dodelson, A.~Grassi, C.~Iossa, D.~Panea~Lichtig, and A.~Zhiboedov, {\it
  {Holographic thermal correlators from supersymmetric instantons}},
  \href{http://arxiv.org/abs/2206.07720}{{\tt arXiv:2206.07720}}.

\bibitem{Bhatta:2022wga}
A.~Bhatta and T.~Mandal, {\it {Exact thermal correlators of holographic
  $CFT$s}},  \href{http://arxiv.org/abs/2211.02449}{{\tt arXiv:2211.02449}}.

\end{thebibliography}\endgroup
\end{document}